\documentclass[12pt]{article}
\usepackage{lscape}
\usepackage{citesort}
\usepackage{amssymb}
\usepackage{appendix}
\usepackage{multirow}

\begin{document}

\def\Tr{\mbox{Tr}}
\def\figt#1#2#3{
        \begin{figure}
        $\left. \right.$
        \vspace*{-2cm}
        \begin{center}
        \includegraphics[width=10cm]{#1}
        \end{center}
        \vspace*{-0.2cm}
        \caption{#3}
        \label{#2}
        \end{figure}
	}
	
\def\figb#1#2#3{
        \begin{figure}
        $\left. \right.$
        \vspace*{-1cm}
        \begin{center}
        \includegraphics[width=10cm]{#1}
        \end{center}
        \vspace*{-0.2cm}
        \caption{#3}
        \label{#2}
        \end{figure}
                }

\def\ds{\displaystyle}
\def\beq{\begin{equation}}
\def\eeq{\end{equation}}
\def\bea{\begin{eqnarray}}
\def\eea{\end{eqnarray}}
\def\beeq{\begin{eqnarray}}
\def\eeeq{\end{eqnarray}}
\def\ve{\vert}
\def\vel{\left|}
\def\ver{\right|}
\def\nnb{\nonumber}
\def\ga{\left(}
\def\dr{\right)}
\def\aga{\left\{}
\def\adr{\right\}}
\def\lla{\left<}
\def\rra{\right>}
\def\rar{\rightarrow}
\def\lrar{\leftrightarrow}  
\def\nnb{\nonumber}
\def\la{\langle}
\def\ra{\rangle}
\def\ba{\begin{array}}
\def\ea{\end{array}}
\def\tr{\mbox{Tr}}
\def\ssp{{\Sigma^{*+}}}
\def\sso{{\Sigma^{*0}}}
\def\ssm{{\Sigma^{*-}}}
\def\xis0{{\Xi^{*0}}}
\def\xism{{\Xi^{*-}}}
\def\qs{\la \bar s s \ra}
\def\qu{\la \bar u u \ra}
\def\qd{\la \bar d d \ra}
\def\qq{\la \bar q q \ra}
\def\GG{\langle g_s^2 G^2 \rangle}
\def\g5{\gamma_5 \not\!q}
\def\x{\gamma_5 \not\!x}
\def\g5{\gamma_5}
\def\sb{S_Q^{cf}}
\def\sd{S_d^{be}}
\def\su{S_u^{ad}}
\def\sbp{{S}_Q^{'cf}}
\def\sdp{{S}_d^{'be}}
\def\sup{{S}_u^{'ad}}
\def\ssp{{S}_s^{'??}}

\def\sig{\sigma_{\mu \nu} \gamma_5 p^\mu q^\nu}
\def\fo{f_0(\frac{s_0}{M^2})}
\def\ffi{f_1(\frac{s_0}{M^2})}
\def\fii{f_2(\frac{s_0}{M^2})}
\def\O{{\cal O}}
\def\sl{{\Sigma^0 \Lambda}}
\def\es{\!\!\! &=& \!\!\!}
\def\ap{\!\!\! &\approx& \!\!\!}
\def\ar{&+& \!\!\!}
\def\arrr{\!\!\!\! &+& \!\!\!}
\def\ek{&-& \!\!\!}
\def\kek{\!\!\!\!&-& \!\!\!}
\def\cp{&\times& \!\!\!}
\def\se{\!\!\! &\simeq& \!\!\!}
\def\eqv{&\equiv& \!\!\!}
\def\kpm{&\pm& \!\!\!}
\def\kmp{&\mp& \!\!\!}
\def\mcdot{\!\cdot\!}
\def\erar{&\rightarrow&}

% .........................................................

\def\simlt{\stackrel{<}{{}_\sim}}
\def\simgt{\stackrel{>}{{}_\sim}}

% .........................................................

\title{
         {\Large
                 {\bf
Octet negative parity to octet positive parity electromagnetic
transitions in light cone QCD
                 }
         }
      }

\author{\vspace{1cm}\\
{\small T. M. Aliev \thanks {
taliev@metu.edu.tr}~\footnote{Permanent address: Institute of
Physics, Baku, Azerbaijan.}\,\,,
M. Savc{\i} \thanks
{savci@metu.edu.tr}} \\
{\small Physics Department, Middle East Technical University,
06531 Ankara, Turkey }}

\date{}

\begin{titlepage}
\maketitle
\thispagestyle{empty}

\begin{abstract}

Light cone QCD sum rules for the electromagnetic transition form factors
among positive and negative parity octet baryons are derived. The unwanted
contributions of the diagonal transitions among  positive  parity octet
baryons are eliminated by combining the sum rules derived from different
Lorentz structures. The $Q^2$ dependence for the transversal and
longitudinal helicity amplitudes are studied.

\end{abstract}

%\vspace{1cm}
~~~PACS numbers: 11.55.Hx, 13.40.Gp, 14.20.Gk

\end{titlepage}

\section{Introduction}

According to the quark-gluon picture baryons are represented as their bound
states, and for this reason what is measured in experiments is actually
indirect manifestation of their realizations. The experimental investigation
for obtaining information about the internal structure of baryons is based
on measurement of the form factors. One main direction in getting
useful information in order to understand the internal structure of baryons
is to study the electromagnetic properties of the baryons. At present,
except the proton and neutron, the electromagnetic form factors of octet
spin-1/2 baryons have not yet been studied experimentally.

The experiments conducted at Jefferson Laboratory (JLab) and Mainz Microton facilty
play the key role in study of the electromagnetic structure of baryons via
the scattering of electrons on nucleons, i.e., $e N \to e N^\ast$, where
$N^\ast$ is the nucleon excitation. These reactions proceed through the
$\gamma^\ast N \to N^\ast$, where $\gamma^\ast$ is the virtual photon, and
these transitions are described by the electromagnetic form factors. The
study of the properties of nucleon excitations, in particular the form
factors constitute one of the main research programs of the
above-mentioned laboratories.

In one of our previous works we analyzed the $\gamma^\ast N \to N^\ast(1535)$
transition form factors \cite{Rsvc01} within the light cone QCD sum rules
method (LCSR) \cite{Rsvc02} (for an application of the LCSR on baryon form
factors, see also \cite{Rsvc03}). The experiments that have been conducted at
JLab for this experiment have collected a lot of data. The future planned
experiments at JLab will allow the chance to study the structure of $N^\ast$
at high photon virtualities up to $Q^2=12~GeV^2$ (see \cite{Rsvc04}). This
transition has already been studied in framework of the covariant quark
model \cite{Rsvc05} and lattice gauge theory \cite{Rsvc06}.

In the present work we extend our previous work for the $\gamma^\ast N \to
N^\ast(1535)$ transition \cite{Rsvc01} to all members of the positive and
negative parity spin-1/2 octet baryons. In this regard we analyze the
$\gamma^\ast \Sigma \to \Sigma^\ast$ and $\gamma^\ast \Xi \to \Xi^\ast$
transitions and calculate their transition
form factors within framework of the LCSR. All these transitions are
customarily denoted as $\gamma^\ast B \to B^\ast$, where $B$ represents the
spin-1/2 positive parity; and $B^\ast$ represents the spin-1/2 negative
parity baryons.

The paper is organized in the following way. In section 2, the sum rules for
the form factors of the $\gamma^\ast B \to B^\ast$ transitions are derived
in LCSR. Section 3 is devoted to the numerical analysis, summary and
conclusions.  

\section{Transition form factors of $\gamma^\ast B \to B^\ast$}

It is well known that in describing 
the $\gamma^\ast B \to B^\ast$ transition, the electromagnetic current
$J_\mu^{el}$ is sandwiched 
between $B$ with momentum $p$ and $B^\ast$ with momentum $p^\prime$,
i.e., $\lla B^\ast (p^\prime) \vel J_\mu^{el} (q) \ver B(p) \rra$.
This matrix element is parametrized in terms of two form factors as:
\bea
\label{esvc01}
\lla B^\ast(p^\prime) \vel J_\mu^{el} (q) \ver B(p) \rra \es
\bar{u}_{B^\ast} (p^\prime) \Bigg[ \Bigg( \gamma_\mu - {\rlap/{q} q_\mu
\over q^2} \Bigg) F_1^\ast (Q^2) \nnb \\
\ek {i \over m_B + m_{B^\ast} }
\sigma_{\mu\nu} q^\nu F_2^\ast (Q^2) \Bigg] \gamma_5 u_B(p)~,
\eea
where $F_1^\ast (Q^2)$; and $ F_2^\ast (Q^2)$ are the Dirac and Pauli
form factors, respectively; $q=p-p^\prime$;
and $Q^2=-q^2$. Appearance of the second term in
$F_1^\ast$ is dictated by the conservation of the electromagnetic current.
In order to derive the LCSR for the form factors $F_1^\ast (Q^2)$ and
$F_2^\ast (Q^2)$ we consider the following vacuum-to-ground state positive
parity baryon correlation function:
\bea
\label{esvc02}
\Pi_\alpha (p,q) \es i \int d^4x e^{i q x}\lla 0 \vel {\cal T} \{\eta(0)
J_\alpha^{el}(x) \} \ver B(p) \rra~.
\eea
Here $\eta$ is the interpolating current of the octet baryon, and
$J_\alpha^{el}(x) = e_u \bar{u} \gamma_\alpha u + e_d \bar{d} \gamma_\alpha d
+  e_s \bar{s} \gamma_\alpha s $ is the electromagnetic current. The general
form of the interpolating currents for light octet baryons are given as:
\bea
\label{esvc03}
\eta_{\Sigma^+} \es 2 \varepsilon^{abc} \sum_{\ell=1}^2 (u^{aT} C A_1^\ell s^b)
A_2^\ell  u^c~, \nnb \\
\eta_{\Sigma^-} \es \eta_{\Sigma^+} (u \to d)~, \nnb \\
\eta_{\Sigma^0} \es \sqrt{2}  \varepsilon^{abc} \sum_{\ell=1}^2 \left[(u^{aT} C
A_1^\ell s^b) A_2^\ell  d^c + (d^{aT} C A_1^\ell s^b) A_2^\ell  u^c \right]~, \nnb \\
\eta_{\Xi^0} \es \eta_{\Sigma^+} (u \lrar s)~, \nnb \\
\eta_{\Xi^-} \es \eta_{\Sigma^-} (d \lrar s)~, 
\eea
where $C$ is the charge conjugation operator; and $A_1^1=I$;
$A_1^2=A_2^1=\gamma_5$; and $A_2^2=\beta$.

We consider now the hadronic transitions involving negative parity baryons.
According to the standard procedure for obtaining sum rules for the
corresponding physical quantities, we substitute in Eq. (\ref{esvc01}) the
total set of negative and positive parity baryon states between the
interpolating and electromagnetic currents. The resulting hadronic
dispersion relations contain contributions from the lowest positive parity
baryon and its negative parity partner. The matrix element of the
interpolating current between the vacuum and one-particle positive parity
baryon states is determined as:
\bea
\label{esvc04}    
\lla 0 \vel \eta \ver B(p) \rra = \lambda_B u_B(p)~,
\eea
where $\lambda_B$ is the residue of the corresponding baryon. Similarly, the
matrix element of the interpolating current between the vacuum and
one-particle negative parity baryon states is defined as:
\bea
\label{esvc05}
\lla 0 \vel \eta \ver B^\ast(p) \rra = \lambda_{B^\ast} \gamma_5
u_{B^\ast}(p)~.
\eea
The hadronic matrix elements $\lla B (p-q) \vel J_\alpha^{el} \ver B(p)
\rra$ and $\lla B^\ast (p-q) \vel J_\alpha^{el} \ver B(p) \rra$ are defined
in terms of the form factors. The second of these matrix elements which is
written in terms of the Dirac and Pauli type form factors is given
in Eq. (\ref{esvc01}). The first matrix element describing the
electromagnetic transition among positive parity baryons can be obtained
from Eq. (\ref{esvc01}) by making the following replacements
$F_{1,2}^\ast(Q^2) \to F_{1,2}(Q^2)$, and then omitting the
$\rlap/{q}q_\alpha/q^2$ terms and replacing $\gamma_5$ with the unit matrix.

Using the equation of motion $(\rlap/{p} - m_B) u_B(p)=0$, the correlation
function can be represented in terms of six independent invariant functions
as:
\bea
\label{esvc06}
\Pi_\alpha((p-q)^2,q^2) \es \Pi_1((p-q)^2,q^2) \gamma_\alpha +
\Pi_2((p-q)^2,q^2) q_\alpha + \Pi_3((p-q)^2,q^2)
q_\alpha \rlap/{q} \nnb \\
\ar \Pi_4((p-q)^2,q^2) p_\alpha +\Pi_5((p-q)^2,q^2) p_\alpha \rlap/{q} +
\Pi_6((p-q)^2,q^2) \gamma_\alpha \rlap/{q}~,
\eea
where all invariant functions depend on $(p-q)^2$ and $q^2$.    

Using the definition of the form factors and residues, and performing
summation over the baryon spin we get the following expressions for the
invariant functions:
\bea          
\label{esvc07}
\Pi_1((p-q)^2, q^2) \es - {\lambda_{B^\ast} (m_{B^\ast}+m_B) \over
m_{B^\ast}^2 - (p-q)^2 } F_1^\ast (q^2) -
{\lambda_{B^\ast} (m_{B^\ast}-m_B) \over
m_{B^\ast}^2 - (p-q)^2 } F_2^\ast (q^2) + \cdots \nnb \\ \nnb \\
%%%
\Pi_2((p-q)^2, q^2) \es {\lambda_{B^\ast} (m_{B^\ast}^2 - m_B^2) \over q^2[m_{B^\ast}^2 -
(p-q)^2] } F_1^\ast (q^2) +
{\lambda_{B^\ast} (m_{B^\ast}-m_B) \over
(m_{B^\ast} + m_B) [m_{B^\ast}^2 -(p-q)^2] } F_2^\ast (q^2) \nnb \\
\ar {\lambda_B \over m_B^2 - (p-q)^2} F_2(q^2) + \cdots \nnb \\ \nnb \\
%%%
\Pi_3((p-q)^2, q^2) \es {\lambda_{B^\ast} (m_{B^\ast} + m_B) \over q^2[m_{B^\ast}^2 -
p^{\prime 2}] } F_1^\ast (q^2) +
{\lambda_{B^\ast} \over (m_{B^\ast}+m_B)[m_{B^\ast}^2 - (p-q)^2] }
F_2^\ast (q^2) \nnb \\
\ek {\lambda_B \over 2 m_B [m_B^2 - (p-q)^2]} F_2(q^2) + \cdots
\eea
Here dots correspond to the contributions of the excited and continuum
states with quantum numbers of $B$ and $B^\ast$. According to the quark
hadron duality these contributions are modeled as perturbative ones
starting on from some threshold $s_0$.

Employing the nucleon interpolating current, we now calculate the correlation function 
from the QCD side. In order to justify the expansion of the product of two
current near the light cone $x^2 \simeq 0$, the external momenta are taken in
deep Eucledian domain. The operator product expansion (OPE) is carried out
over twist which involves the distribution amplitudes (DAs) of the baryon
with growing twist. The matrix element of the three quark operators between
the vacuum and the state of the members of the positive parity octet baryons
is defined in terms of the DAs of the baryons, i.e.,
\bea
\label{esvc08}
\varepsilon^{abc} \lla 0 \vel q_{1\alpha}^a (a_1x) q_{2\beta}^b (a_2x)
q_{3\gamma}^c (a_3x) \ver B(p) \rra~,
\eea
where $a,b,c$ are the color indices; $a_1,a_2$; and $a_3$ are positive
numbers. 

Using the Lorentz covariance, and parity and spin of the baryons this matrix
element can be written in terms of 27 DAs as:
\bea
\label{esvc09} 
4 \varepsilon^{abc} \lla 0 \vel q_{1\alpha}^a (a_1x) q_{2\beta}^b (a_2x)    
q_{3\gamma}^c (a_3x) \ver B(p) \rra = \sum_{i} {\cal F}_i
\Gamma_{\alpha\beta}^{1i} \left[\Gamma^{2i} B(p) \right]_\gamma~,
\eea
where $\Gamma^i$ are certain Dirac matrices; and ${\cal F}_i$ are the DAs
which do not posses definite twist. For completeness, the matrix element
(\ref{esvc09}) is presented in Appendix A.  

The matrix element given in Eq. (\ref{esvc08}) is defined in terms of the
definite twist DAs as:
\bea
\label{nolabel01}
4 \varepsilon^{abc} \lla 0 \vel q_{1\alpha}^a (a_1x) q_{2\beta}^b (a_2x)    
q_{3\gamma}^c (a_3x) \ver B(p) \rra = \sum_{i} F_i
\Gamma_{\alpha\beta}^{\prime 1i} \left[\Gamma^{\prime 2i} B(p) \right]_\gamma~, \nnb
\eea
and the two sets of DAs are connected to each other by the following
relations:
\bea
\label{esvc10}
\begin{array}{ll}
{\cal S}_1 = S_1~,& (2 P \mcdot x) \, {\cal S}_2 = S_1 - S_2~,  \\
{\cal P}_1 = P_1~,& (2 P \mcdot x) \, {\cal P}_2 = P_2 - P_1~,  \\
{\cal V}_1 = V_1~,& (2 P \mcdot x) \, {\cal V}_2 = V_1 - V_2 - V_3~,  \\
2{\cal V}_3 = V_3~,& (4 P \mcdot x) \, {\cal V}_4 =
- 2 V_1 + V_3 + V_4 + 2 V_5~, \\
(4 P\mcdot x) \, {\cal V}_5 = V_4 - V_3~,&
(2 P \mcdot x)^2 \, {\cal V}_6 = - V_1 + V_2 + V_3 + V_4 + V_5 - V_6~,  \\
{\cal A}_1 = A_1~,& (2 P \mcdot x) \, {\cal A}_2 = - A_1 + A_2 - A_3~,  \\
2 {\cal A}_3 = A_3~,&
(4 P \mcdot x) \, {\cal A}_4 = - 2 A_1 - A_3 - A_4 + 2 A_5~,  \\
(4 P \mcdot x) \, {\cal A}_5 = A_3 - A_4~,&
(2 P \mcdot x)^2 \, {\cal A}_6 = A_1 - A_2 + A_3 + A_4 - A_5 + A_6~,  \\
{\cal T}_1 = T_1~, & (2 P \mcdot x) \, {\cal T}_2 = T_1 + T_2 - 2T_3~,  \\
2 {\cal T}_3 = T_7~,& (2 P \mcdot x) \, {\cal T}_4 = T_1 - T_2 - 2 T_7~,  \\
(2 P\mcdot x) \, {\cal T}_5 = - T_1 + T_5 + 2 T_8~,&
(2 P \mcdot x)^2 \, {\cal T}_6 = 2 T_2 - 2 T_3 - 2 T_4 + 2 T_5 + 2 T_7 + 2
T_8~, \\
(4P \mcdot x) \, {\cal T}_7 = T_7 - T_8~, &
(2 P\mcdot x)^2 \, {\cal T}_8 = - T_1 + T_2 + T_5 - T_6 + 2 T_7 + 2 T_8~.
\end{array}
\eea

We also present the explicit expressions of the DAs with definite twist in
Appendix B.

The calculation of the invariant amplitudes from the QCD side is tedious but
straightforward. The invariant amplitudes in terms of the spectral densities
$\rho_{2i}$, $\rho_{4i}$, and $\rho_{6i}$ can be written as:
\bea
\label{esvc11}
\Pi_i = N \int_0^1 dx \Bigg\{ {\rho_{2i}(x) \over (q-px)^2} +
{\rho_{4i}(x) \over (q-px)^4} +
{\rho_{6i}(x) \over (q-px)^6} \Bigg\}~,
\eea
where $N=2$ for the $\Sigma^+$, $\Sigma^-$, $\Xi^0$, and  $\Xi^-$; and
$\sqrt{2}$ for $\Sigma^0$ baryons.
Explicit expressions of the spectral densities $\rho_{2i}$, $\rho_{4i}$, and 
$\rho_{6i}$ are presented in Appendix C.
It should be noted that the sum rules derived from combinations of different
Lorentz structures are suggested in \cite{Rsvc07}.

Equating Eqs. (\ref{esvc07}) and (\ref{esvc11}), and performing Borel
transformation over $-(p-q)^2$ we get the following sum rules for the
$\gamma^\ast B \to B^\ast$ transition form factors:
\bea
\label{esvc12} 
\lambda_{B^\ast} \left[m_{B^\ast}+m_B) F_1^\ast(Q^2) + m_{B^\ast}-m_B)
F_2^\ast(Q^2)\right] e^{-m_{B^\ast}^2/M^2} \es -I_1(Q^2,M^2,s_0)~, \nnb \\
\lambda_{B^\ast} \left[ - {m_{B^\ast}+m_B)\over Q^2} F_1^\ast(Q^2) +
F_2^\ast(Q^2)\right] e^{-m_{B^\ast}^2/M^2} \es -I_1(Q^2,M^2,s_0) \nnb \\
\ar 2 m_B I_3(Q^2,M^2,s_0)~,
\eea
and $I_i(Q^2,M^2,s_0)$ are determined to be:
\bea
\label{esvc13}
I_i(Q^2,M^2,s_0) \es
\int_{x_0}^1 dx \Bigg[
- {\rho_{2i}(x)\over x} + {\rho_{4i}(x) \over x^2 M^2} -
- {\rho_{6i}(x) \over 2 x^3 M^4} \Bigg] e^{-s(x)/M^2} \nnb \\
\ar \Bigg[ {\rho_{4i}(x_0) \over Q^2 + x_0^2 m_B^2} -
{1\over 2 x_0}
{\rho_{6i}(x_0) \over (Q^2 + x_0^2 m_B^2) M^2} \nnb \\
\ar {1\over 2} {x_0^2 \over (Q^2 + x_0^2 m_B^2)} \Bigg(
{d\over dx_0} {\rho_{6i}(x_0) \over x_0 (Q^2 + x_0^2 m_B^2)
M^2} \Bigg) \Bigg]e^{-s_0/ M^2}
\Bigg\}~,
\eea
where \bea
\label{nolabel02}
s(x) = {\bar{x} Q^2 + x \bar{x} m_B^2 \over x}~, \nnb
\eea
and $x_0$ is the solution of $s(x) = s_0$.

As has already been noted, the analysis of the experimental data for the
nucleon system is performed with the Dirac and Pauli type form factors. The
data for the nucleon resonances is mostly analyzed with the help of helicity
amplitudes (see for example \cite{Rsvc04,Rsvc08,Rsvc09}). In other words, in the
analysis of data related to the negative parity baryons it is more suitable
to study the helicity amplitudes instead of Dirac and Pauli type form
factors. The electro production of negative parity spin-1/2 baryon resonance
in the $\gamma^\ast B \to B^\ast$ transitions is described with the help of
two independent, transverse amplitude $A_{1/2}$ and longitudinal amplitude
$S_{1/2}$. The relations among the form factors $F_1^\ast$ and $F_2^\ast$,
and helicity amplitudes $A_{1/2}$  and $S_{1/2}$ is given as:
\bea
\label{esvc14}
A_{1/2} \es - 2 e \sqrt{ {(m_{B^\ast} + m_B)^2 +Q^2 \over 8 m_B (m_{B^\ast}^2
- m_B^2) } } \Bigg[ F_1^\ast (Q^2) + {m_{B^\ast} - m_B \over m_{B^\ast} + m_B}
F_2^\ast (Q^2) \Bigg]~, \\ \nnb \\
\label{egsn15}
S_{1/2} \es \sqrt{2} e \sqrt{ {(m_{B^\ast} + m_B)^2 +Q^2 \over 8 m_B
(m_{B^\ast}^2 - m_B^2) } }
{\vel \vec{q} \ver \over Q^2}
\Bigg[ {m_{B^\ast} - m_B \over m_{B^\ast} + m_B}  F_1^\ast (Q^2) \nnb \\
\ek  {Q^2 \over (m_{B^\ast} + m_B)^2 } F_2^\ast (Q^2) \Bigg]~,
\eea
where $e$ is the electric charge; and $\vec{q}$ is the photon
three-momentum whose absolute value in the rest frame of $B^\ast$ is given as:
\bea
\label{nolabel03}
\vel \vec{q} \ver ={\sqrt{ Q^4 + 2 Q^2(m_{B^\ast}^2+m_B^2) +
(m_{B^\ast}^2 - m_B^2)^2} \over 2 m_{B^\ast}}~. \nnb
\eea

In determining the form factors $F_1^\ast$ and $F_2^\ast$ from the sum
rules given in Eq. (\ref{esvc12}) the residues $\lambda_{B^\ast}$ of the
negative parity baryons are needed, which is obtained from the two-point
correlation function
\bea
\label{nolabel04}
\Pi(p^2) = i \int d^4x e^{ipx} \lla 0 \vel \mbox{T} \left\{ \eta(x)
\bar{\eta}(0) \right\} \ver 0 \rra~.\nnb
\eea
Following the same procedure presented in \cite{Rsvc01}, we get for the mass
and residue of the negative parity spin-1/2 octet baryons
\bea
\label{esvc16}
m_{B^\ast}^2 \es { \ds \int_0^{s_0} ds \, e^{-s/M^2} s \Big[ m_B \mbox{Im}
\Pi_1(s) - \mbox{Im} \Pi_2 (s) \Big] \over
\ds \int_0^{s_0} ds \, e^{-s/M^2} \Big[ m_B \mbox{Im}
\Pi_1(s) - \mbox{Im} \Pi_2 (s) \Big] }~, \\ \nnb \\
\label{esvc17}
\vel \lambda_{B^\ast} \ver^2 \es {e^{m_{B^\ast}^2/M^2} \over m_{B^\ast} + m_B}
{1\over \pi} \int_0^{s_0} ds \,e^{-s/M^2} \Big[ m_B \mbox{Im}
\Pi_1(s) - \mbox{Im} \Pi_2 (s) \Big]~,
\eea  
$\mbox{Im}\Pi_1(s)$ and $\mbox{Im} \Pi_2 (s)$ correspond to the spectral
densities for the $\rlap/{p}$ and unit operator structures, respectively,
and they are calculated in \cite{Rsvc10}.

\section{Numerical analysis}
In the previous section we have calculated the $\gamma^\ast B \to B^\ast$
transition form factors and helicity amplitudes using within the framework
of the LCSR method. In this section we will present our numerical results on
the helicity amplitudes.

The main nonperturbative contributions to LCSR are realized by the DAs,
which are presented in the Appendix. In the numerical analysis we will use
the DAs of the $\Sigma$, $\Xi$ and $\lambda$ baryons which are calculated in
\cite{Rsvc11,Rsvc12,Rsvc13}. The parameters appearing in the expressions of
the DAs are determined from the two-point QCD sum rules, and their values
are given as:
\bea
\label{esvc18}
f_\Xi \es (9.9 \pm 0.4)\times 10^{-3}~GeV^2~, \nnb \\
\lambda_1 \es -(2.1 \pm 0.1)\times 10^{-2}~GeV^2~, \nnb \\
\lambda_2 \es (5.2 \pm 0.2)\times 10^{-2}~GeV^2~, \nnb \\
\lambda_3 \es (1.7 \pm 0.1)\times 10^{-2}~GeV^2~, \nnb \\ \nnb \\
f_\Sigma \es (9.4 \pm 0.4)\times 10^{-3}~GeV^2~, \nnb \\
\lambda_1 \es -(2.5 \pm 0.1)\times 10^{-2}~GeV^2~, \nnb \\
\lambda_2 \es (4.4 \pm 0.1)\times 10^{-2}~GeV^2~, \nnb \\
\lambda_3 \es (2.0 \pm 0.1)\times 10^{-2}~GeV^2~,
\eea
We have recalculated these parameters once more and obtained that the sum
rules of these parameters given in \cite{Rsvc09,Rsvc10,Rsvc11} contain
errors. Our reanalysis on these parameters predicts that:
\bea
\label{esvc19}
f_\Xi \es (11.70 \pm 0.4)\times 10^{-3}~GeV^2~, \nnb \\
\lambda_1 \es -(3.15 \pm 0.1)\times 10^{-2}~GeV^2~, \nnb \\
\lambda_2 \es (6.50 \pm 0.2)\times 10^{-2}~GeV^2~, \nnb \\
\lambda_3 \es (2.15 \pm 0.1)\times 10^{-2}~GeV^2~, \nnb \\ \nnb \\
f_\Sigma \es (13.00 \pm 0.4)\times 10^{-3}~GeV^2~, \nnb \\
\lambda_1 \es -(3.15 \pm 0.1)\times 10^{-2}~GeV^2~, \nnb \\
\lambda_2 \es (6.75 \pm 0.1)\times 10^{-2}~GeV^2~, \nnb \\
\lambda_3 \es (1.80 \pm 0.1)\times 10^{-2}~GeV^2~,
\eea
which we shall use in our numerical analysis. The masses of the negative
parity baryons are taken from the QCD sum rules estimation give in Eq.
(\ref{esvc17}) having the values: $m_{\Sigma^\ast}=(1.7\pm0.1)~GeV$
$m_{\Xi^\ast}=(1.75\pm0.1)~GeV$, and $m_{\Lambda^\ast}=(1.7\pm0.1)~GeV$
which are very close to the experimental results. For the quark condensate
we use $\qq (1~GeV) = -\left(246_{-19}^{+28}~MeV\right)^3$ \cite{Rsvc14}.

The domain of the Borel mass parameter used in the calculations for the form
factors is chosen to be $M^2=(1.8\pm 0.4)~GeV^2$, which is decided with the
criteria that the power and continuum contributions are sufficiently
suppressed. The value of the continuum threshold is determined in such a way
that the mass sum rules prediction reproduce the experimentally measured
mass to within the limits of 10-15\% accuracy, and this condition leads that
$s_0=(3.7\pm 0.3)~GeV^2$. The working region of the arbitrary parameter
$\beta$ is determined from the condition that $\lambda_{B^\ast}^2$ be
positive and exhibit good stability with respect to the variation in
$\beta$. Our analysis shows that the residues of all negative parity baryons  
satisfy the above-required conditions in the range $0.4 \le \beta \le 0.8$,
which we shall use in further numerical analysis.

In Figs. (1) and (2) we depict the photon momentum square $Q^2$ dependence
of the helicity amplitudes $A_{1/2}$ and $S_{1/2}$ for the $\gamma^\ast
\Sigma^+ \to \Sigma^{+\ast}$, respectively, at
$M^2=1.6~GeV^2$, $s_0=3.5~GeV^2$, and at three fixed values of $\beta$
picked from its working region. For the parameters $\lambda_1$, $\lambda_2$
and $\lambda_3$ appearing in DAs, we use our own results given in Eq.
(\ref{esvc19}). In order to keep higher twist, continuum and higher states
contributions under control $Q^2$ is restricted vary in the domain
$1~GeV^2 \le Q^2 \le 10~GeV^2$. 
It follows from Fig. (1) that, $A_{1/2}$ decreases with increasing $Q^2$ and
tends to zero asymptotically. The situation for $S_{1/2}$ is presented in
Fig. (2), from which we observe that it also mimics the behavior of
$A_{1/2}$ and tends to zero at large $Q^2$. We see that the transversal
helicity amplitude is 3 to 4 times smaller in modulo compared to the
longitudinal helicity amplitude $S_{1/2}$ at all values of $Q^2$.

In Figs. (3) and (4) the dependencies of $A_{1/2}$ and $S_{1/2}$ on $Q^2$ at
the same values of $M^2$ and $s_0$ are presented for the $\gamma^\ast
\Sigma^- \to \Sigma^{-\ast}$ transition, respectively. The trends in regard
to their dependence on $Q^2$ are same, i.e., both amplitudes decrease with
increasing $Q^2$ in modulo. We also observe that the values of the modulo of
$A_{1/2}$ and $S_{1/2}$ are small compared to the $\gamma^\ast\Sigma^+ \to
\Sigma^{+\ast}$ transition, at least 2 to 3 times.

In Figs. (5) and (6) we present the $Q^2$ dependence of the transversal and
longitudinal helicity amplitudes for the $\gamma^\ast\Sigma^0 \to
\Sigma^{0\ast}$ transition. We see from these figures that the magnitude of
$A_{1/2}$ seems to be slightly smaller compared to the $\gamma^\ast \Sigma^-
\to \Sigma^{-\ast}$ case, while the magnitude of $S_{1/2}$ appears to be 
approximately 50\% larger compared to the same transition.

The $Q^2$ dependence of $A_{1/2}$ and $S_{1/2}$ for the $\gamma^\ast \Xi^-
\to \Xi^{-\ast}$ transition are given in Figs. (7) and (8). We observe from
these figures that the values of $A_{1/2}$ are quite similar to the ones
predicted for the $\gamma^\ast \Sigma^- \to \Sigma^{-\ast}$ transition. In
the case of $S_{1/2}$ however, the difference between the transitions is around
40\%.

Finally, Figs. (9) and (10) depict the dependence of the helicity amplitudes
on $Q^2$ for the $\gamma^\ast \Xi^0 \to \Xi^{0\ast}$ transition. It follows
from these figures that, $A_{1/2}$ change its sign at $Q^2=1.5~GeV^2$ at
the fixed value of the arbitrary parameter $\beta=0.8$. The maximum value
$A_{1/2}$ is equal to $0.04$ at $Q^2=1~GeV^2$, when $\beta=0.4$. We further
see that the magnitude of $S_{1/2}$ is quite close to the one predicted for
the $\gamma^\ast \Sigma^0 \to \Sigma^{0\ast}$ transition.

We can summarize our results as follows:

\begin{itemize}

\item The transversal helicity amplitude $A_{1/2}$ seems to be practically
insensitive to the values of the arbitrary parameter $\beta$ for the
$\gamma^\ast \Sigma^- \to \Sigma^{-\ast}$ and $\gamma^\ast \Xi^- \to
\Xi^{-\ast}$ transitions. 

\item Contrary to the above behavior, the same amplitude $A_{1/2}$ for the
$\gamma^\ast \Sigma^+ \to \Sigma^{+\ast}$ is quite sensitive to the value of
$\beta$. The value of $A_{1/2}$ at $Q^2=1~GeV^2$ doubles itself when $\beta$
changes from $0.4$ to $0.8$.

\item The longitudinal amplitude $S_{1/2}$ does weakly depend on $\beta$ for
all considered transitions.

\end{itemize}

Of course measurement of these electromagnetic form factors is quite
difficult due to the short life-time of hyperons. We hope that along with
further developments in experimental techniques, measurement of these
transition form factors could become possible.  

It should be noted here that, our results can be improved further with the
help of more reliable calculations of DAs and with the inclusion of
perturbative ${\cal O}(\alpha_s)$ corrections, and the first attempt in this
direction has already been made in \cite{Rsvc15}.

In conclusion, we investigate the electromagnetic transition among
octet positive and negative parity baryons within LCSR method. We calculate
the transversal and longitudinal helicity amplitudes described by these
transitions. The $Q^2$ dependence of these amplitudes are studied. We show
that the longitudinal helicity amplitude seems to be practically insensitive
to the variations in the arbitrary parameter $\beta$ for all considered
transitions.

\newpage

\section*{Appendix A}
%\section*{}
\setcounter{equation}{0}
In this appendix we present the general Lorentz decomposition of
the matrix element of the three-quark operators between the vacuum and the
octet baryon states in terms of the DAs \cite{Rsvc16}.

\bea
\label{appendixA}
\lefteqn{ 4 \lla 0 \vel \varepsilon^{ijk} u_\alpha^i(a_1 x) u_\beta^j(a_2 x) d_\gamma^k(a_3 x)
\ver B(p) \rra = } \nnb \\
&&  
{\cal S}_1 m_B C_{\alpha \beta} \left(\gamma_5 B\right)_\gamma + 
{\cal S}_2 m_B^2 C_{\alpha \beta} \left(\!\not\!{x} \gamma_5 B\right)_\gamma + 
{\cal P}_1 m_B \left(\gamma_5 C\right)_{\alpha \beta} B_\gamma + 
{\cal P}_2 m_B^2 \left(\gamma_5 C \right)_{\alpha \beta} \left(\!\not\!{x} B\right)_\gamma 
\nnb \\
\ar 
\left({\cal V}_1+\frac{x^2m_B^2}{4}{\cal V}_1^M \right)
\left(\!\not\!{p}C \right)_{\alpha \beta} \left(\gamma_5 B\right)_\gamma + 
{\cal V}_2 m_B \left(\!\not\!{p} C \right)_{\alpha \beta} \left(\!\not\!{x} \gamma_5 B\right)_\gamma  + 
{\cal V}_3 m_B  \left(\gamma_\mu C \right)_{\alpha \beta}\left(\gamma^{\mu} \gamma_5 B\right)_\gamma 
\nnb \\ 
\ar
{\cal V}_4 m_B^2 \left(\!\not\!{x}C \right)_{\alpha \beta} \left(\gamma_5 B\right)_\gamma +
{\cal V}_5 m_B^2 \left(\gamma_\mu C \right)_{\alpha \beta} \left(i \sigma^{\mu\nu} x_\nu \gamma_5 
B\right)_\gamma 
+ {\cal V}_6 m_B^3 \left(\!\not\!{x} C \right)_{\alpha \beta} \left(\!\not\!{x} \gamma_5 B\right)_\gamma  
\nnb \\ 
\ar 
\left({\cal A}_1+\frac{x^2m_B^2}{4}{\cal A}_1^M\right)
\left(\!\not\!{p}\gamma_5 C \right)_{\alpha \beta} B_\gamma + 
{\cal A}_2 m_B \left(\!\not\!{p}\gamma_5 C \right)_{\alpha \beta} \left(\!\not\!{x} B\right)_\gamma  + 
{\cal A}_3 m_B  \left(\gamma_\mu \gamma_5 C \right)_{\alpha \beta}\left( \gamma^{\mu} B\right)_\gamma 
\nnb \\ 
\ar
{\cal A}_4 m_B^2 \left(\!\not\!{x} \gamma_5 C \right)_{\alpha \beta} B_\gamma +
{\cal A}_5 m_B^2 \left(\gamma_\mu \gamma_5 C \right)_{\alpha \beta} \left(i \sigma^{\mu\nu} x_\nu  
B\right)_\gamma 
+ {\cal A}_6 m_B^3 \left(\!\not\!{x} \gamma_5 C \right)_{\alpha \beta} \left(\!\not\!{x} B\right)_\gamma  
\nnb \\
\ar
\left({\cal T}_1+\frac{x^2m_B^2}{4}{\cal T}_1^M\right)
\left(p^\nu i \sigma_{\mu\nu} C\right)_{\alpha \beta} \left(\gamma^\mu\gamma_5 B\right)_\gamma 
+ 
{\cal T}_2 m_B \left(x^\mu p^\nu i \sigma_{\mu\nu} C\right)_{\alpha \beta} \left(\gamma_5 B\right)_\gamma 
\nnb \\
\ar 
 {\cal T}_3 m_B \left(\sigma_{\mu\nu} C\right)_{\alpha \beta} \left(\sigma^{\mu\nu}\gamma_5 B\right)_\gamma 
+ {\cal T}_4 m_B \left(p^\nu \sigma_{\mu\nu} C\right)_{\alpha\beta} \left(\sigma^{\mu\rho} x_\rho \gamma_5 B\right)_\gamma 
\nnb \\
\ar {\cal T}_5 m_B^2 \left(x^\nu i \sigma_{\mu\nu} C\right)_{\alpha \beta} \left(\gamma^\mu\gamma_5 B\right)_\gamma 
+ 
{\cal T}_6 m_B^2 \left(x^\mu p^\nu i \sigma_{\mu\nu} C\right)_{\alpha \beta} \left(\!\not\!{x} \gamma_5 B\right)_\gamma  
\nnb \\
\ar {\cal T}_{7} m_B^2 \left(\sigma_{\mu\nu} C\right)_{\alpha \beta} \left(\sigma^{\mu\nu} \!\not\!{x} \gamma_5 B\right)_\gamma
%\nnb \\&&
+ {\cal T}_{8} m_B^3 \left(x^\nu \sigma_{\mu\nu} C\right)_{\alpha \beta} \left(\sigma^{\mu\rho} x_\rho \gamma_5 B\right)_\gamma ~,
\nnb
\eea
where $C$ is the charge conjugation operator; and $B$ represents the
octet baryon with momentum $p$. 

\newpage

\section*{Appendix B}  
\setcounter{equation}{0}
\bea
\label{nolabel05}
\mbox{\bf twist-4 DAs:} \nnb \\ \nnb \\
V_2(x_i)\es 24 x_1 x_2 [\phi_4^0 + \phi_4^+ (1- 5 x_3)]~,\nnb \\
A_2(x_i) \es 24 x_1 x_2 (x_2 -x_1) \phi_4^-~, \nnb \\
T_2(x_i) \es24 x_1 x_2 [ \xi_4^{0} + \xi_4^+ (1- 5 x_3) ]~, \nnb\\
V_3(x_i) \es 12 x_3 \{\psi_4^0 (1-x_3) + \psi_4^+(1-x_3 -10 x_1
x_2)+ \psi_4^-[x_1^2+x_2^2-x_3 (1-x_3)] \} ~, \nnb \\
A_3(x_i) \es 12 x_3 (x_2-x_1) [ (\psi_4^0+\psi_4^+) +
\psi_4^- (1-2 x_3)]~, \nnb \\
T_3(x_i)\es6 x_3  \{B(\phi_4^0 + \psi_4^0 + \xi_4^0 )(1-x_3)
+(\phi_4^+ + \psi_4^+ + \xi_4^+ ) (1-x_3 -10 x_1 x_2) \nnb \\
\ar (\phi_4^- - \psi_4^-+ \xi_4^- ) [x_1^2 + x_2^2 - x_3 (1-x_3)]  \}~, \nnb \\
T_7(x_i)\es 6 x_3 \{ (\phi_4^0 + \psi_4^0 - \xi_4^0 )(1-x_3)
+(\phi_4^+ + \psi_4^+ - \xi_4^+ ) (1-x_3 -10 x_1 x_2)  \nnb \\
\ar (\phi_4^- - \psi_4^- - \xi_4^- ) [x_1^2 + x_2^2 - x_3 (1-x_3)] \}~, \nnb\\
S_1(x_i)\es6 x_3 (x_2 -x_1) [( \phi_4^0 + \psi_4^0 + \xi_4^0 +\phi_4^+ + \psi_4^+ + \xi_4^+ )
+ (\phi_4^- - \psi_4^- + \xi_4^- ) (1- 2 x_3)]~,\nnb \\
P_1(x_i)\es 6 x_3 (x_1 -x_2) [( \phi_4^0 + \psi_4^0 - \xi_4^0 +\phi_4^+ + \psi_4^+ - \xi_4^+ )
+ (\phi_4^- - \psi_4^- - \xi_4^- ) (1- 2 x_3)]~, \nnb\\ \nnb \\
\mbox{\bf twist-5 DAs:} \nnb \\ \nnb \\
V_4(x_i)\es 3 \{\psi_5^0 (1-x_3) + \psi_5^+ [1-  x_3 - 2 (x_1^2 +x_2^2)]
+ \psi_5^- [2 x_1 x_2 -x_3 (1-x_3)] \} ~,\nnb\\
A_4(x_i)\es3  (x_2 -x_1) [ -\psi_5^{0} + \psi_5^+(1- 2 x_3) + \psi_5^- x_3]~,\nnb \\
T_4(x_i)\es\frac{3}{2} \{ (\phi_5^{0} + \psi_5^{0} + \xi_5^0 ) (1 - x_3) + 
(\phi_5^+ + \psi_5^+ + \xi_5^+ )  [1-  x_3 - 2 (x_1^2 + x_2^2)] \} \nnb \\
\ar (\phi_5^- - \psi_5^- + \xi_5^- )   [2 x_1 x_2 - x_3 (1- x_3)] ~, \nnb\\
T_8(x_i)\es\frac{3}{2}  \{ ( \phi_5^0 + \psi_5^{0} - \xi_5^0 )
(1-x_3) + (\phi_5^+ +\psi_5^+-\xi_5^+ ) [1-x_3 -2 (x_1^2 + x_2^2)] \nnb \\
\ar (\phi_5^- - \psi_5^- + \xi_5^- )  [2 x_1 x_2 - x_3 (1- x_3)] \}~,\nnb \\
V_5(x_i)\es 6 x_3 [\phi_5^0 + \phi_5^+ (1-2 x_3)  ]~, \nnb \\
A_5(x_i)\es 6 x_3 (x_2 - x_1) \phi_5^- ~,\nnb \\
T_5(x_i)\es6 x_3 [  \xi_5^0 + \xi_5^+  (1- 2 x_3) ]~, \nnb\\
S_2(x_i)\es \frac{3}{2} (x_2 -x_1) [-( \phi_5^0 + \psi_5^0 + \xi_5^0 )
+ (\phi_5^+ + \psi_5^+ + \xi_5^+ ) (1- 2 x_3) \nnb \\
\ar (\phi_5^- - \psi_5^- + \xi_5^- ) x_3]~, \nnb  \\
P_2(x_i)\es \frac{3}{2} (x_1 -x_2) [-( \phi_5^0 + \psi_5^0 - \xi_5^0 )
+ (\phi_5^+ + \psi_5^+ - \xi_5^+ ) (1- 2 x_3) \nnb \\
\ar (\phi_5^- - \psi_5^- - \xi_5^- ) x_3]~, \nnb \\ \nnb \\
\mbox{\bf twist-6 DAs:} \nnb \\ \nnb \\
V_6(x_i)\es2 [  \phi_6^0  + \phi_6^+ (1- 3 x_3) ]~, \nnb \\
A_6(x_i)\es2 (x_2-x_1)\phi_6^- ~,\nnb \\
T_6(x_i)\es2 [ \phi_6^0  - {1 \over 2}(\phi_6^+-\phi_6^-) (1- 3 x_3)~.\nnb
\eea

\newpage

\section*{Appendix C}
%\section*{}
\setcounter{equation}{0}	

In this appendix we present the expressions for the functions
$\rho_{2i}$, $\rho_{4i}$ and $\rho_{6i}$
which appear in the sum rules for $F_i^\ast(Q^2)$
for the $\gamma^\ast
\Sigma^+ \to \Sigma^{\ast +}$ transition.

%\section*{$\gamma^\ast \Sigma^+ \to \Sigma^{\ast +}$ transition}
\bea
%\Sigma^{\ast +}
\label{Sigma+}
%F1-rho6
%
\rho_{61}^{\Sigma^{\ast +}} (x)\es
%e_u
e_u m_B^3 Q^2 {(Q^2+m_B^2 x^2)\over x} 
\Big[4 (m_{B^\ast}-m_B) (1+\beta) (2-x)+ 
m_u (1-\beta) x\Big] \;\check{\!\check{B}}_6 (x) \nnb \\
%e_u
\ar e_u m_u m_B^2 Q^2 \Big\{ 4 m_B^2 (m_{B^\ast}-m_B)
(1+\beta)(2-x) \Big(\; \widetilde{\!\widetilde{C}}_6 + \;
\widetilde{\!\widetilde{D}}_6 \Big) \nnb \\
\ar {(1-\beta)\over x} \Big[ m_B^2 [ 8 m_{B^\ast} - 
m_B( 8-x^2) ] x \; \widetilde{\!\widetilde{B}}_6  
+ Q^2 [ 4 m_{B^\ast} - 
m_B( 4-x) ] \; \widetilde{\!\widetilde{B}}_6 \nnb \\
\ek 8 m_B^2 (m_{B^\ast}-m_B) x (2-x)
\; \widetilde{\!\widetilde{B}}_8 \Big]\Big\} (x) \nnb \\
%e_s
\ar e_s  m_B^2 {Q^2 \over x} \Big\{ 4 m_s m_B^2 
(m_{B^\ast}-m_B) (1-\beta) (2-x) x \Big(\;
\widehat{\!\widehat{C}}_6 - \; \widehat{\!\widehat{D}}_6 \Big) \nnb \\
\ar 4 (1+\beta) m_B (m_{B^\ast}-m_B) (2-x)  \Big[ 2 m_s m_B x
\; \widehat{\!\widehat{B}}_8 +(Q^2+m_B^2 x^2) \;
\widehat{\!\widehat{B}}_6 \Big] \nnb \\
\ar (1+\beta) m_s \Big[ x m_B^2
[8 m_{B^\ast} - m_B (8 - x^2) ] +
Q^2 [4 m_{B^\ast} - m_B (4 - x) ] \Big]
\;\widehat{\!\widehat{B}}_6 \Big\} (x) \nnb \\ \nnb \\
\rho_{41}^{\Sigma^{\ast +}} (x)\es
%e_u
{1\over 2} e_u m_B^3 Q^2 (1-\beta) \Big\{ 2 \left[ 2 m_{B^\ast} -
m_B (2-x)\right] \Big(\;\check{\!\check{C}}_6 +
\;\check{\!\check{D}}_6 \Big) + m_u \;\check{\!\check{B}}_6 \Big\} (x) \nnb \\
\ek e_u m_B^3 {Q^2\over x} (1+\beta) \Big\{ \Big[ m_B [ 8 -
(5-x) x ] - m_{B^\ast} (8 - 5 x) \Big] \;\check{\!\check{B}}_6  \nnb \\
\ek 3 x [ 2 m_{B^\ast} - m_B (2-x) ] \;\check{\!\check{B}}_8 \Big\} (x) \nnb \\
\ar e_u m_B^2 {Q^2 \over 2 x} (1-\beta) \Big\{ 2 m_B x 
[ 2 m_{B^\ast} - m_B (2-x)] \Big(\;
\widetilde{\!\widetilde{C}}_6 - \; \widetilde{\!\widetilde{D}}_6 \Big) \nnb \\
\ek m_u \Big[ m_B (8+x) \; \widetilde{\!\widetilde{B}}_6
- 8 m_{B^\ast} \; \widetilde{\!\widetilde{B}}_6 - 4 m_B x
 \; \widetilde{\!\widetilde{B}}_8 \Big] \Big\} (x)\nnb \\
\ar e_u  m_B^2 {Q^2 \over 2 x} (1+\beta) \Big\{
Q^2 \; \widetilde{\!\widetilde{B}}_6 - m_B x [m_B x + 4
(m_{B^\ast} - m_B)] \Big(\;
\widetilde{\!\widetilde{B}}_6 - 2 \; \widetilde{\!\widetilde{B}}_8 \Big) \nnb \\
\ek 2 m_u m_B x \Big(\; \widetilde{\!\widetilde{C}}_6 +
 \; \widetilde{\!\widetilde{D}}_6 \Big) \Big\} (x) \nnb \\
\ek e_s  m_B^3 Q^2 (1-\beta) \Big[ m_B x
\Big(\;\widehat{\!\widehat{C}}_6 + \; \widehat{\!\widehat{D}}_6
\Big) + m_s \Big(\;\widehat{\!\widehat{C}}_6 - \;
\widehat{\!\widehat{D}}_6 \Big)\Big] (x) \nnb \\
\ar e_s m_B^2 {Q^2 \over 2 x} (1+\beta) \Big\{ 2 m_B x [2
(m_s + m_{B^\ast}) - m_B (2+x)] \; \widehat{\!\widehat{B}}_8 \nnb \\
\ar \Big[2 m_B ( m_{B^\ast} - m_B)
(8-3 x) - 2 Q^2 + m_s [8 m_{B^\ast} - m_B (8+x)] \Big]\; \widehat{\!\widehat{B}}_6
\Big\}(x) \nnb \\
\ek e_u m_B {Q^2 \over 2 x} (1-\beta) \Big\{
2 m_B (x^3 m_B^2 - 2 Q^2) \Big(\check{C}_2 + \check{D}_2 \Big) \nnb \\
\ar 2 m_B x \Big[ 2 m_B (m_{B^\ast}-m_B) 
\Big( 2 \check{C}_2 + \check{C}_4 -
3 \check{C}_5 + 2 \check{D}_2 - \check{D}_4 + 3 \check{D}_5 \Big)
+ Q^2 \Big(\check{C}_2 + \check{D}_2 \Big) \Big] \nnb \\
\ar Q^2 \Big[ 4 m_{B^\ast} \Big(\check{C}_2 + \check{D}_2 \Big) +
m_u \Big(\check{B}_2 + 5 \check{B}_4 \Big) \Big] \nnb \\
\ek m_B^2 x^2 \Big[2 (m_{B^\ast}-m_B) \Big(
\check{C}_4 - 3 \check{C}_5 - \check{D}_4 + 3 \check{D}_5 \Big) \nnb \\
\ar  m_u \Big(  \check{B}_2 -
\check{B}_4 + 6 \check{B}_5 + 12 \check{B}_7 -
2 \check{E}_1 + 2 \check{H}_1\Big) \Big] \Big\} (x) \nnb \\
\ar e_u m_B {Q^2 \over 2 x} (1+\beta) \Big\{
-[2 (m_{B^\ast}-m_B) Q^2 + m_B^3 x^3 ] 
\Big( \check{B}_2 + 5 \check{B}_4 \Big) \nnb \\
\ar 2 m_B^2 (m_{B^\ast}-m_B) x^2 \Big(
\check{B}_2 - \check{B}_4 + 6 \check{B}_5 + 12 \check{B}_7 -
2 \check{E}_1 + 2 \check{H}_1\Big) \nnb \\
\ar m_B x \Big[
-Q^2 \Big(\check{B}_2 + 5 \check{B}_4 \Big) -
8 m_B (m_{B^\ast}-m_B) \Big(
\check{B}_2 + 2 \check{B}_4 + 3 \check{B}_5 + 6 \check{B}_7 -
\check{E}_1 + \check{H}_1\Big)\nnb \\
\ar m_u \Big[ m_B^2 x^2 \Big( \check{C}_4 - 3 \check{C}_5
- \check{B}_4 + 3 \check{B}_5 \Big) - Q^2 \Big( \check{C}_2 +
\check{D}_2 \Big) \Big] \Big\} (x) \nnb \\
\ek e_u m_B {Q^2 \over 2 x} (1-\beta) \Big\{
2 m_B^2 (m_{B^\ast}-m_B) (2-x) x
\Big(\widetilde{C}_4 - \widetilde{C}_5 + \widetilde{D}_4 -
\widetilde{D}_5 \Big) \nnb \\
\ar m_u \Big[ Q^2 \Big( \widetilde{B}_2 + \widetilde{B}_4 \Big)
+ 4 m_B (m_{B^\ast}-m_B) x
\Big( \widetilde{B}_2 - \widetilde{B}_4 + 2 \widetilde{B}_5\Big) \nnb \\
\ek m_B^2 x \Big( \widetilde{B}_2 - \widetilde{B}_4 + 2
\widetilde{B}_5 + 2 \widetilde{E}_1 + 2 \widetilde{H}_1 \Big) \Big]
\Big\} (x) \nnb \\
\ar e_u m_B {Q^2 \over 2 x} (1+\beta) \Big\{
- [4 (m_{B^\ast}-m_B) Q^2 + m_B^3 x^3]
\Big( \widetilde{B}_2 + \widetilde{B}_4 \Big) \nnb \\
\ar 8 m_B^2 (m_{B^\ast}-m_B) x^2 
\Big( \widetilde{B}_5 + 2 \widetilde{B}_7 \Big) +
m_B x \Big[ -Q^2 \Big( \widetilde{B}_2 + \widetilde{B}_4 \Big) \nnb \\
\ek  8 m_B (m_{B^\ast}-m_B) 
\Big( \widetilde{B}_2 + \widetilde{B}_4 + 2
\widetilde{B}_5 + 4 \widetilde{B}_7 \Big) \Big] \nnb \\
\ar m_u \Big[ - m_B x [2 m_{B^\ast} - m_B (2 -x)] 
\Big( \widetilde{C}_4 - \widetilde{D}_4 \Big) \nnb \\  
\ar m_B x [2 m_{B^\ast} - m_B (2 +x)]
\Big( \widetilde{C}_5 -  \widetilde{D}_5 \Big)
+ 2 Q^2 \Big( \widetilde{C}_2 +  \widetilde{D}_2 \Big) \Big] \Big\} (x) \nnb \\
\ek e_s m_B {Q^2 \over 2 x} (1-\beta) \Big\{ 4 (m_{{\cal
O}^\ast}-m_B) (2 m_B^2 x + Q^2) \Big( \widehat{C}_2 +
\widehat{D}_2 \Big) \nnb \\
\ek 4 m_B^2 (m_{B^\ast} - m_B)
(2-x) x \Big( \widehat{C}_5 - \widehat{D}_5 \Big)
+ m_s \Big[ - 2  Q^2  \Big( \widehat{C}_2 -
\widehat{D}_2 \Big) \nnb \\ 
\ar  m_B x [2 m_{B^\ast} - m_B (2
-x)] \Big( \widehat{C}_4 + \widehat{D}_4 \Big) +
m_B x [2 m_{B^\ast} - m_B (2  
+x)] \Big( \widehat{C}_5 + \widehat{D}_5 \Big) \Big] \Big\} (x) \nnb \\
\ar e_s m_B {Q^2 \over 2 x} (1+\beta) \Big\{   
2 (m_{B^\ast}-m_B) Q^2 \Big( \widehat{B}_2 - 3 \widehat{B}_4
\Big) + m_B^3 x^3 \Big( \widehat{B}_2 + \widehat{B}_4\Big) \nnb \\
\ar 2 m_B^2 (m_{B^\ast}-m_B) x^2
\Big( \widehat{B}_2 - \widehat{B}_4 + 2 \widehat{B}_5 +
4 \widehat{B}_7 - 2 \widehat{E}_1 + 2 \widehat{H}_1 \Big) \nnb \\
\ar m_B x \Big[ Q^2 \Big( \widehat{B}_2 + \widehat{B}_4\Big)
- 8 m_B (m_{B^\ast}-m_B) \Big( \widehat{B}_4 + \widehat{B}_5 +
2 \widehat{B}_7 - 2 \widehat{E}_1 + \widehat{H}_1 \Big) \Big] \nnb \\
\ek m_s \Big[ Q^2 \Big( \widehat{B}_2 + \widehat{B}_4\Big) -
m_B^2 x^2 \Big( \widehat{B}_2 - \widehat{B}_4 +
2 \widehat{B}_5 - 2 \widehat{E}_1 - 2 \widehat{H}_1 \Big) \nnb \\
\ek 4 m_B (m_{B^\ast}-m_B) x \Big( \widehat{B}_2 -
\widehat{B}_4 + 2 \widehat{B}_5 \Big) \Big] \Big\} (x) \nnb \\
\ek e_u m_B^3 [ 2 m_{B^\ast} - m_B (2-x)] Q^2
\int_0^{\bar{x}} dx_3 \, \Big[(1-\beta) (A_1^M - V_1^M) - 3 (1+\beta) T_1^M
\Big](x,1-x-x_3,x_3) \nnb \\
\ar e_u m_B^2 {Q^2\over x}
\int_0^{\bar{x}} dx_1 \, \Big\{ (1-\beta) Q^2 (A_1^M+V_1^M) \nnb \\
\ar m_B x (1+\beta) [4 m_{B^\ast} - m_B (4-x)]
T_1^M \Big\}(x_1,x,1-x_1-x) \nnb \\
\ar e_s m_B^2 Q^2\int_0^{\bar{x}} dx_1 \, \Big\{ [2 m_B ( m_B - m_{B^\ast})
x - Q^2] (1-\beta) (A_1^M - V_1^M) \nnb \\
\ar  m_B x [2 m_{B^\ast} - m_B
(2+x)] (1+\beta) T_1^M \Big\}(x_1,1-x_1-x,x) \nnb \\ \nnb \\
%F1-rho6
%
\rho_{21}^{\Sigma^{\ast +}} (x)\es
%e_u
e_u m_B^2 {Q^2 \over 2 x} (1+\beta) \;
\widetilde{\!\widetilde{B}}_6 (x)\nnb \\
\ek e_s m_B^2 {Q^2 \over x} (1+\beta) \;                       
\widehat{\!\widehat{B}}_6 (x)\nnb \\
\ek e_u m_B {Q^2\over 2 x} \Big\{(1+\beta) \Big[
(m_{B^\ast} - m_B) \Big( \check{B}_2 + 5\check{B}_4 \Big)
 + m_u \Big( \check{C}_2 + \check{D}_2 \Big) \Big]\nnb \\
\ar (1-\beta) \Big[ 4 (m_{B^\ast} - m_B)
\Big( \check{B}_2 + \check{B}_4 \Big) + m_u \Big( \check{B}_2 +
5\check{B}_4 \Big) \Big] \Big\} (x) \nnb \\
\ar e_u m_B {Q^2\over 2 x} \Big\{2 (1+\beta) \Big[
-2 (m_{B^\ast} - m_B) \Big( \widetilde{B}_2 + \widetilde{B}_4 \Big) \nnb \\
\ar m_B x \Big( \widetilde{B}_4 + \widetilde{B}_5 + 2 \widetilde{B}_7
+ \widetilde{E}_1 - \widetilde{H}_1 \Big) +m_u \Big( \widetilde{C}_2 +
\widetilde{D}_2 \Big) \Big] \nnb \\
\ar (1-\beta) \Big[ 2 m_B x \Big( \widetilde{C}_2 - \widetilde{C}_5 -
\widetilde{D}_2 - \widetilde{D}_5 \Big) - m_u \Big( \widetilde{B}_2 +  
\widetilde{B}_4 \Big) \Big] \Big\} (x) \nnb \\
\ek e_s m_B {Q^2\over 2 x} \Big\{(1-\beta) \Big[
4 (m_{B^\ast} - m_B) \Big( \widehat{C}_2 - \widehat{D}_2 \Big)
- m_B x \Big( \widehat{C}_4 - \widehat{C}_5 - \widehat{D}_4 +
\widehat{D}_5 \Big) \nnb \\
\ek 2 m_s \Big( \widehat{C}_2 - \widehat{D}_2 \Big) \Big]
- (1+\beta) \Big[ 2 m_{B^\ast} \Big( \widehat{B}_2 - 3
\widehat{B}_4 \Big) - 2 m_B (1-x) \widehat{B}_2 +
6 m_B \widehat{B}_4 \nnb \\
\ar 2 m_B  x \Big( \widehat{B}_4 +
2 \widehat{B}_5 + 4 \widehat{B}_7 \Big) -
m_s \Big( \widehat{B}_2 + \widehat{B}_4 \Big) \Big] \Big\} (x) \nnb \\
\ar e_u m_B Q^2
\{[2 m_{B^\ast} - m_B (2-x)]
(1+\beta) + m_u (1-\beta)\} \nnb \\
\cp \int_0^{\bar{x}}dx_3 \, (P_1 + S_1 + 3 T_1 - 6 T_3)
(x,1-x-x_3,x_3) \nnb \\
\ek e_u  m_B Q^2 \{[2 m_{B^\ast} - m_B (2-x)]
(1-\beta) + m_u (1+\beta)\} \nnb \\
\cp \int_0^{\bar{x}} dx_3 \, 
( A_1 + 2 A_3 - V_1 + 2 V_3 ) (x,1-x-x_3,x_3) \nnb \\
\ar e_u m_B Q^2 (1+\beta) \int_0^{\bar{x}} dx_1 \, 
\Big[4 (m_{B^\ast} - m_B)(T_1-2 T_3) +
m_B x (P_1 + S_1 +T_1 - 2 T_3) \nnb \\
\ar m_u (A_1 + A_3 - V_1 + V_3) \Big] (x_1,x,1-x_1-x) \nnb \\
\ar e_u m_B {Q^2 \over x} (1-\beta) \int_0^{\bar{x}} dx_1 \,
\Big[ m_B^2 A_1^M + 2 m_B (m_{B^\ast} - m_B) x
(A_3-V_3) + Q^2 (A_1+V_1) \nnb \\
\ar m_u m_B x (P_1-S_1+T_1) \Big]
(x_1,x,1-x_1-x) \nnb \\
\ek e_s m_B Q^2 (1+\beta) \int_0^{\bar{x}} dx_1 \, \Big\{
m_B \Big[ (2-x) (P_1+S_1) + (2+x) (T_1-2 T_3) \Big] \nnb \\
\ek 2 m_{B^\ast} (P_1 + S_1 +T_1 - 2 T_3)
+ m_s (P_1 - S_1 - T_1) \Big\} (x_1,1-x_1-x,x) \nnb \\
\ek e_s {Q^2 \over x}(1-\beta) \int_0^{\bar{x}} dx_1 \, \Big\{  
m_B x \Big[2 m_{B^\ast} (A_1 + A_3 - V_1 + V_3) +
m_s (A_1 + A_3 + V_1 -V_3) \Big] \nnb \\
\ek  Q^2 (A_1-V_1) - m_B^2 \Big[(A_1^M - V_1^M) + 2 x
(A_1+A_3-V_1+V_3) \Big] \Big\}(x_1,1-x_1-x,x) \nnb \\ \nnb \\
\rho_{62}^{\Sigma^{\ast +}} (x)\es
%e_u
-e_u m_B^3 {(Q^2+m_B^2 x^2)\over x}    
[4 Q^2 (2-x) (1+\beta) - m_u (m_{B^\ast}+m_B) x
(1-\beta)] \;\check{\!\check{B}}_6 (x) \nnb \\
\ek e_u m_u m_B^2 {1\over x} \Big\{ 4 m_B^2 Q^2 (2-x) 
x (1+\beta) \Big(\; \widetilde{\!\widetilde{C}}_6 + \;
\widetilde{\!\widetilde{D}}_6 \Big) \nnb \\
\ek (1-\beta) [ m_B^3
(m_{B^\ast}+m_B) x^3 + m_B (m_{B^\ast} -
 7 m_B) Q^2 x - 4 Q^4 ] \; \widetilde{\!\widetilde{B}}_6 \nnb \\
\ar 8 m_B^2 Q^2 (2-x) x (1-\beta) \; \widetilde{\!\widetilde{B}}_8 \Big\} (x) \nnb \\
\ek 4 e_s m_s m_B^4 Q^2 (2-x) (1-\beta) \Big(\;
\widehat{\!\widehat{C}}_6 - \; \widehat{\!\widehat{D}}_6 \Big) (x) \nnb \\
\ek e_s m_B^2 {1\over x} (1+\beta) \Big\{
4 m_B Q^2 (Q^2+m_B^2 x^2)
(2-x) \;\widehat{\!\widehat{B}}_6 \nnb \\
\ek m_s [ m_B^3 
(m_{B^\ast}+m_B) x^3 + m_B (m_{B^\ast} -        
 7 m_B) Q^2 x - 4 Q^4 ] \; \widehat{\!\widehat{B}}_6 \nnb \\
\ek 8 m_s m_B^2 Q^2 (2-x) x \; \widehat{\!\widehat{B}}_8
\Big\} (x) \nnb \\ \nnb \\
\rho_{42}^{\Sigma^{\ast +}} (x)\es
- e_u  m_B^3 {1\over 2 x} \Big\{ 2 (1+\beta) \Big[
8 Q^2 \;\check{\!\check{B}}_6 - x Q^2 \Big(
5\;\check{\!\check{B}}_6 - 6 \;\check{\!\check{B}}_8\Big) +
m_B (m_{B^\ast}+m_B) x^2 \Big(
\;\check{\!\check{B}}_6 - 3 \;\check{\!\check{B}}_8\Big) \Big] \nnb \\
\ar  x (1-\beta) \Big[
2 [m_B (m_{B^\ast}+m_B) x - 2 Q^2]
\Big( \;\check{\!\check{C}}_6 + \;\check{\!\check{D}}_6\Big) -
m_u (m_{B^\ast}+m_B) \;\check{\!\check{C}}_6 \Big] \Big\} (x) \nnb \\
\ar e_u  m_B^2 {1\over 2 x} \Big\{ (1+\beta) \Big[
(m_{B^\ast}+m_B) Q^2 \; \widetilde{\!\widetilde{B}}_6 -
m_B [m_B (m_{B^\ast}+m_B) x - 4 Q^2] x \Big( \;
\widetilde{\!\widetilde{B}}_6 - 2 \; \widetilde{\!\widetilde{B}}_8 \Big) \nnb \\
\ek 2 m_u m_B (m_{B^\ast}+m_B) x \Big( \;
\widetilde{\!\widetilde{C}}_6 + \; \widetilde{\!\widetilde{D}}_6 \Big)\Big] \nnb \\
\ar (1-\beta) \Big[ 2 m_B [m_B (m_{B^\ast}+m_B) x - 2 Q^2] x
\Big( \; \widetilde{\!\widetilde{C}}_6 - \; \widetilde{\!\widetilde{D}}_6 \Big) \nnb \\
\ek 8 m_u Q^2 \; \widetilde{\!\widetilde{B}}_6 - m_u m_B
(m_{B^\ast}+m_B) x \Big( \;
\widetilde{\!\widetilde{B}}_6 - 4 \; \widetilde{\!\widetilde{B}}_8
\Big)\Big] \Big\} (x)\nnb \\
\ek e_s  m_B^3 (m_{B^\ast}+m_B) {1\over 2 x} \Big\{
2 x (1-\beta) \Big[ m_B x \Big( \; \widehat{\!\widehat{C}}_6 +
\; \widehat{\!\widehat{D}}_6 \Big) + m_s \Big( \; \widehat{\!\widehat{C}}_6
- \; \widehat{\!\widehat{D}}_6 \Big) \Big] \nnb \\
\ar (1+\beta) \Big[ 2 m_B [m_B (m_{B^\ast}+m_B) x
+ 2 Q^2 - 2 m_s (m_{B^\ast}+m_B)] x \; \widehat{\!\widehat{B}}_8 \nnb \\ 
\ar 2 Q^2 [m_{B^\ast} + 3 m_B (3-x)] \;
\widehat{\!\widehat{B}}_6 + m_s [m_B (m_{B^\ast}+m_B)
x + 8 Q^2 ] \; \widehat{\!\widehat{B}}_6 \Big] \Big\} (x) \nnb \\
\ek e_u  m_B {1\over 2 x} (1-\beta) \Big\{
2 m_B^3 x^3 (m_{B^\ast}+m_B) \check{C}_2 +
2 m_B^2 Q^2 x^2 \Big(\check{C}_4 - 3 \check{C}_5 -
\check{D}_4 + 3 \check{D}_5 \Big) \nnb \\
\ek 2 m_B Q^2 x \Big[m_B \Big( 3 \check{C}_2 + 2\check{C}_4 
- 6 \check{C}_5 - 2 \check{D}_4 + 6 \check{D}_5\Big) -
m_{B^\ast} \check{C}_2 \Big] \nnb \\
\ek 4 Q^4 \check{C}_2 +
2 [m_B^3 x^3 (m_{B^\ast}+m_B)  + m_B x
(m_{B^\ast}- 3 m_B) Q^2 - 2 Q^4] \check{D}_2 \nnb \\
\ar m_u (m_{B^\ast}+m_B) \Big[ Q^2 \Big(
\check{B}_2 + 5 \check{B}_4 \Big) \nnb \\
\ek m_B^2 x^2 \Big(
\check{B}_2 - \check{B}_4 + 6 \check{B}_5 + 12 \check{B}_7 - 2 \check{E}_1
+ 2 \check{H}_1 \Big) \Big] \Big\} (x) \nnb \\
\ar e_u  m_B {1\over 2 x} (1+\beta) \Big\{
[2 Q^4 - m_B^3 (m_{B^\ast}+m_B) x^3] \Big(  
\check{B}_2 + 5 \check{B}_4 \Big) \nnb \\
\ek 2 m_B^2 Q^2 x^2 \Big(
\check{B}_2 - \check{B}_4 + 6 \check{B}_5 + 12 \check{B}_7 - 2 \check{E}_1
+ 2 \check{H}_1 \Big) - m_B Q^2 x \Big[ m_{B^\ast} \Big(
\check{B}_2 + 5 \check{B}_4 \Big) \nnb \\
\ek m_B
\Big( 7 \check{B}_2 + 11 \check{B}_4 + 24 \check{B}_5 + 48 \check{B}_7 - 8
\check{E}_1  + 8 \check{H}_1 \Big) \Big]\nnb \\
\ar m_u (m_{B^\ast}+m_B) \Big[m_B^2
x^2 \Big(\check{C}_4 - 3 \check{C}_5 - \check{D}_4 + 3 \check{D}_5 \Big) - 2 Q^2
\Big( \check{C}_2 + \check{D}_2 \Big) \Big] \Big\} (x)\nnb \\
\ar e_u  m_B {1\over 2 x} (1-\beta) \Big\{
2 m_B^2 Q^2 (2-x) x \Big(\widetilde{C}_4 - \widetilde{C}_5 +
\widetilde{D}_4 - \widetilde{D}_5 \Big) \nnb \\  
\ek m_u \Big[ (m_{B^\ast}+m_B) Q^2 \Big(\widetilde{B}_2 +
\widetilde{B}_4\Big) + 4 m_B Q^2 x \Big(\widetilde{B}_2 - 
\widetilde{B}_4 + 2 \widetilde{B}_5\Big) \nnb \\
\ek m_B^2 x^2 (m_{B^\ast}+m_B) \Big(\widetilde{B}_2 - 
\widetilde{B}_4 + 2 \widetilde{B}_5  +2 \widetilde{E}_1 + 2
\widetilde{H}_1 \Big) \Big] \Big\} (x) \nnb \\
\ar e_u  m_B {1\over 2 x} (1+\beta) \Big\{
[4 Q^4 - m_B^3 (m_{B^\ast}+m_B) x^3] \Big(  
\widetilde{B}_2 + \widetilde{B}_4 \Big) \nnb \\
\ek 8 m_B^2 Q^2 x^2 \Big(  
\widetilde{B}_5 + 2 \widetilde{B}_7 \Big) -
m_B Q^2 x \Big[ m_{B^\ast} \Big(\widetilde{B}_2 +
\widetilde{B}_4 \Big) \nnb \\
\ek m_B \Big(
7 \widetilde{B}_2 + 7 \widetilde{B}_4 + 16 \widetilde{B}_5 + 32
\widetilde{B}_7\Big) \Big] -
m_u \Big[ m_B^2 (m_{B^\ast}+m_B) x^2 \Big(
\widetilde{C}_4 - \widetilde{C}_5 - \widetilde{D}_4 + \widetilde{D}_5 \Big) \nnb \\
\ek 2 m_B Q^2 x \Big(\widetilde{C}_4 + \widetilde{C}_5 -
\widetilde{D}_4 - \widetilde{D}_5 \Big) - 2 (m_{B^\ast}+m_B) Q^2
\Big(\widetilde{C}_2 + \widetilde{D}_2 \Big) \Big] \Big\} (x) \nnb \\
\ar e_s  m_B {1\over 2 x} (1-\beta) \Big\{
4 m_B^2 Q^2 x \Big[ 2 \Big( \widehat{C}_2 + \widehat{D}_2 \Big)
- (2-x) \Big( \widehat{C}_5 -\widehat{D}_5 \Big) \Big] \nnb \\
\ar 4 Q^4 \Big( \widehat{C}_2 +\widehat{D}_2 \Big) -
m_s \Big[ m_B^2 x^2 (m_{B^\ast}+m_B)
\Big( \widehat{C}_4 - \widehat{C}_5 +\widehat{D}_4 - \widehat{D}_5\Big) \nnb \\
\ek 2 m_B Q^2 x \Big(\widehat{C}_4 + 
\widehat{C}_5 +\widehat{D}_4 + \widehat{D}_5\Big) +
(m_{B^\ast}+m_B) Q^2 \Big(\widehat{C}_2 - \widehat{D}_2 \Big)
\Big]\Big\} (x) \nnb \\
\ar e_s  m_B {1\over 2 x} (1+\beta) \Big\{
m_B^3 (m_{B^\ast}+m_B) x^3 \Big( \widehat{B}_2 +
\widehat{B}_4 \Big) - 2 Q^4
\Big( \widehat{B}_2 - 3 \widehat{B}_4 \Big) \nnb \\
\ek 2  m_B^2 Q^2 x^2 \Big(
\widehat{B}_2 - \widehat{B}_4 + 2 \widehat{B}_5 + 4 \widehat{B}_7 - 2 \widehat{E}_1
+ 2 \widehat{H}_1 \Big) \nnb \\
\ar m_B Q^2 x \Big[ m_{B^\ast} \Big( \widehat{B}_2 + \widehat{B}_4
\Big) + m_B \Big( \widehat{B}_2 + 9 \widehat{B}_4 + 8 \widehat{B}_5 + 16
\widehat{B}_7 - 8 \widehat{E}_1 + 8 \widehat{H}_1 \Big) \Big] \nnb \\
\ek m_s \Big[(m_{B^\ast}+m_B) Q^2 \Big( \widehat{B}_2 +
\widehat{B}_4 \Big) - m_B^2 (m_{B^\ast}+m_B) x^2
\Big( \widehat{B}_2 - \widehat{B}_4 + 2 \widehat{B}_5 - 2 \widehat{E}_1 
- 2 \widehat{H}_1 \Big) \nnb \\
\ar 4 m_B Q^2 x \Big( \widehat{B}_2 - \widehat{B}_4 +
2 \widehat{B}_5 \Big) \Big] \Big\} (x) \nnb \\
\ek e_u m_B^3 [ m_B (m_{B^\ast}+m_B) x - 2 Q^2]
\int_0^{\bar{x}} dx_3 \, \Big[ (1-\beta) (A_1^M-V_1^M) \nnb \\
\ek 3 (1+\beta) T_1^M
\Big] (x,1-x-x_3,x_3) \nnb \\
\ar e_u m_B^2 {1\over x}
\int_0^{\bar{x}} dx_1 \, \Big\{ (1-\beta) (m_{B^\ast}+m_B)
Q^2 (A_1^M+V_1^M) \nnb \\
\ar (1+\beta) m_B [m_B (m_{B^\ast} - m_B) x -
4 Q^2] x T_1^M \Big\}(x_1,x,1-x_1-x) \nnb \\
\ar e_s m_B^2 {1\over x} \int_0^{\bar{x}} dx_1 \, \Big\{ 
(1-\beta) [m_{B^\ast}+m_B (1+2 x)] Q^2 (A_1^M - V_1^M) \nnb \\  
\ek (1+\beta)  m_B [m_B (m_{B^\ast}+m_B) x
+ 2 Q^2] x  T_1^M \Big\} (x_1,1-x_1-x,x) \nnb \\ \nnb \\
\rho_{22}^{\Sigma^{\ast +}} (x)\es
e_u  m_B^2 (m_{B^\ast}+m_B) {1\over 2 x} (1+\beta)
\;\widetilde{\!\widetilde{B}}_6 (x) \nnb \\
\ek e_s  m_B^2 (m_{B^\ast}+m_B) {1\over 2 x} (1+\beta)
\;\widehat{\!\widehat{B}}_6 (x) \nnb \\
\ar e_u  m_B {1\over 2 x} \Big\{ 2 (1+\beta) \Big[
Q^2 \Big( \check{B}_2 + 5 \check{B}_4 \Big) -
m_u (m_{B^\ast}+m_B) \Big( \check{C}_2 + \check{D}_2 \Big)
\Big] \nnb \\ 
\ar (1-\beta) \Big[ 4 Q^2 \Big(\check{C}_2 + \check{D}_2 \Big) -
m_u (m_{B^\ast}+m_B) \Big( \check{B}_2 + 5 \check{B}_4
\Big) \Big] \Big\} (x) \nnb \\
\ar e_u  m_B {1\over 2 x} \Big\{ 2 (1+\beta)
\Big[ 2 Q^2 \Big( \widetilde{B}_2 + \widetilde{B}_4 \Big) 
+ m_B (m_{B^\ast}+m_B) x \Big(
\widetilde{B}_4 + \widetilde{B}_5 + 2 \widetilde{B}_7 + \widetilde{E}_1
- \widetilde{H}_1 \Big) \nnb \\
\ar m_u (m_{B^\ast}+m_B)
\Big(\widetilde{C}_2 + \widetilde{D}_2 \Big) \Big]
+ (m_{B^\ast}+m_B) (1-\beta) \Big[
2 m_B x \Big(\widetilde{C}_2 - \widetilde{C}_5 -
\widetilde{D}_2 - \widetilde{D}_5\Big) \nnb \\
\ek  m_u \Big(\widetilde{B}_2 + \widetilde{B}_4 \Big) \Big] \Big\} (x) \nnb \\
\ar e_s  m_B {1\over 2 x} \Big\{ (1-\beta)
\Big[ m_B (m_{B^\ast}+m_B) x \Big(
\widehat{C}_4 - \widehat{C}_5 -  \widehat{D}_4 + \widehat{D}_5
\Big) + 4 Q^2 \Big( \widehat{C}_2 +  \widehat{D}_2 \Big) \nnb \\
\ar 2 m_s (m_{B^\ast}+m_B) \Big( \widehat{C}_2 -
\widehat{D}_2 \Big) \Big] \nnb \\
\ek (1+\beta) \Big[2 Q^2 \Big( \widehat{B}_2 - 3 \widehat{B}_4 \Big) 
- 2  m_B (m_{B^\ast}+m_B) x \Big(
\widehat{B}_2 + \widehat{B}_4 + 2 \widehat{B}_5 + 4 \widehat{B}_7  
\Big) \nnb \\
\ar m_s (m_{B^\ast}+m_B) \Big( \widehat{B}_2 +
\widehat{B}_4 \Big) \Big] \Big\} (x) \nnb \\
\ar e_u m_B (1+\beta)\int_0^{\bar{x}} dx_3 \,
\Big\{[m_B (m_{B^\ast}+m_B) x - 2 Q^2]
(P_1 + S_1 + 3 T_1 - 6 T_3) \nnb \\
\ek m_u (m_{B^\ast}+m_B) 
 (A_1+2 A_3 - V_1 + 2 V_3) \Big\} (x,1-x-x_3,x_3) \nnb \\
\ek e_u m_B (1-\beta)\int_0^{\bar{x}} dx_3 \, \Big\{
[m_B (m_{B^\ast}+m_B) x - 2 Q^2]
(A_1 + 2 A_3 - V_1 + 2 V_3) \nnb \\
\ar  m_u (m_{B^\ast}+m_B)
(P_1 + S_1 + 3 T_1 - 6 T_3) \Big\} (x,1-x-x_3,x_3) \nnb \\
\ek e_u m_B (1+\beta) \int_0^{\bar{x}} dx_1 \, \Big[
4 Q^2 (T_1-2 T_3) - m_B (m_{B^\ast}+m_B) x
(P_1+S_1+T_1-2 T_3) \nnb \\
\ek m_u (m_{B^\ast}+m_B) (A_1+A_3-V_1+V_3) \Big] 
(x_1,x,1-x_1-x) \nnb \\
\ek e_u {1\over x} (1-\beta) \int_0^{\bar{x}} dx_1 \, \Big\{
- m_B^2 (m_{B^\ast}+m_B) (A_1^M+V_1^M) \nnb \\
\ek Q^2 \Big[ (m_{B^\ast}+m_B) (A_1+V_1) -
2 m_B x (A_3-V_3) \Big] \nnb \\
\ek  m_u m_B (m_{B^\ast}+m_B) x (P_1-S_1+T_1)
\Big\} (x_1,x,1-x_1-x) \nnb \\
\ek e_s m_B (1+\beta) \int_0^{\bar{x}} dx_1 \, \Big[
2 Q^2 (P_1+S_1+T_1-2 T_3) - m_B (m_{B^\ast}+m_B) x
(P_1+S_1-T_1+2 T_3) \nnb \\
\ar m_s (m_{B^\ast}+m_B) (P_1-S_1-T_1)
\Big] (x_1,1-x_1-x,x) \nnb \\
\ek e_s m_B (1+\beta) \int_0^{\bar{x}} dx_1 \, \Big[
2 Q^2 (P_1+S_1+T_1-2 T_3) - m_B (m_{B^\ast}+m_B) x
(P_1+S_1-T_1+2 T_3) \nnb \\
\ar m_s (m_{B^\ast}+m_B) (P_1-S_1-T_1)
\Big] (x_1,1-x_1-x,x) \nnb \\
\ar e_s m_B {1\over x} (1-\beta) \int_0^{\bar{x}} dx_1 \, \Big\{
m_B^2 (m_{B^\ast}+m_B) (A_1^M-V_1^M) + 
Q^2 \Big[ (m_{B^\ast}+m_B + 2 x m_B) (A_1-V_1) \nnb \\
\ar 2 m_B x (A_3+V_3) \Big] -
m_s  m_B (m_{B^\ast}+m_B) x (A_1+A_3+V_1-V_3)
 \Big\} (x_1,1-x_1-x,x) \nnb  
\eea

In the above expressions for $\rho_{2i}$, $\rho_{4i}$ and $\rho_{6i}$
the functions ${\cal F}(x_i)$ are defined in the following way:

\bea
\label{nolabel06}
\check{\cal F}(x_1) \es \int_1^{x_1}\!\!dx_1^{'}\int_0^{1- x^{'}_{1}}\!\!dx_3\,
{\cal F}(x_1^{'},1-x_1^{'}-x_3,x_3)~, \nnb \\
\check{\!\!\!\;\check{\cal F}}(x_1) \es 
\int_1^{x_1}\!\!dx_1^{'}\int_1^{x^{'}_{1}}\!\!dx_1^{''}
\int_0^{1- x^{''}_{1}}\!\!dx_3\,
{\cal F}(x_1^{''},1-x_1^{''}-x_3,x_3)~, \nnb \\
%%%%
\widetilde{\cal F}(x_2) \es \int_1^{x_2}\!\!dx_2^{'}\int_0^{1- x^{'}_{2}}\!\!dx_1\,
{\cal F}(x_1,x_2^{'},1-x_1-x_2^{'})~, \nnb \\
\widetilde{\!\widetilde{\cal F}}(x_2) \es 
\int_1^{x_2}\!\!dx_2^{'}\int_1^{x^{'}_{2}}\!\!dx_2^{''}
\int_0^{1- x^{''}_{2}}\!\!dx_1\,
{\cal F}(x_1,x_2^{''},1-x_1-x_2^{''})~, \nnb \\
%%%%
\widehat{\cal F}(x_3) \es \int_1^{x_3}\!\!dx_3^{'}\int_0^{1- x^{'}_{3}}\!\!dx_1\,
{\cal F}(x_1,1-x_1-x_3^{'},x_3^{'})~, \nnb \\
\widehat{\!\widehat{\cal F}}(x_3) \es 
\int_1^{x_3}\!\!dx_3^{'}\int_1^{x^{'}_{3}}\!\!dx_3^{''}
\int_0^{1- x^{''}_{3}}\!\!dx_1\,
{\cal F}(x_1,1-x_1-x_3^{''},x_3^{''})~.\nnb
\eea

Definitions of the functions $B_i$, $C_i$, $D_i$, $E_1$ and $H_1$
that appear in the expressions for $\rho_i(x)$ are given as follows:

\bea
\label{nolabel07}
B_2 \es T_1+T_2-2 T_3~, \nnb \\
B_4 \es T_1-T_2-2 T_7~, \nnb \\
B_5 \es - T_1+T_5+2 T_8~, \nnb \\
B_6 \es 2 T_1-2 T_3-2 T_4+2 T_5+2 T_7+2 T_8~, \nnb \\
B_7 \es T_7-T_8~, \nnb \\
B_8 \es  -T_1+T_2+T_5-T_6+2 T_7+2T_8~, \nnb \\
C_2 \es V_1-V_2-V_3~, \nnb \\
C_4 \es -2V_1+V_3+V_4+2V_5~, \nnb \\
C_5 \es V_4-V_3~, \nnb \\
C_6 \es -V_1+V_2+V_3+V_4+V_5-V_6~, \nnb \\
D_2 \es -A_1+A_2-A_3~, \nnb \\
D_4 \es -2A_1-A_3-A_4+2A_5~, \nnb \\
D_5 \es A_3-A_4~, \nnb \\
D_6 \es A_1-A_2+A_3+A_4-A_5+A_6~, \nnb \\
E_1 \es S_1-S_2~, \nnb \\
H_1 \es P_2-P_1~. \nnb
\eea

\newpage

\section*{Figure captions}
{\bf Fig. 1} The dependence of the helicity amplitude $A_{1/2}$
for the $\gamma^\ast \Sigma^+ \to \Sigma^{+\ast}$ transition
on $Q^2$ at $M^2=1.6~\mbox{GeV}^2$, $s_0=3.5~\mbox{GeV}^2$, and at
several fixed values of the auxiliary parameter $\beta$. \\ \\
{\bf Fig. 2} The same as in Fig. 1, but for the helicity amplitude
$S_{1/2}$. \\ \\
{\bf Fig. 3} The same as in Fig. 1, but for the $\gamma^\ast \Sigma^- \to
\Sigma^{-\ast}$ transition. \\ \\
{\bf Fig. 4} The same as in Fig. 3, but for the helicity amplitude
$S_{1/2}$. \\ \\
{\bf Fig. 5} The same as in Fig. 1, but for the $\gamma^\ast \Sigma^0 \to
\Sigma^{0\ast}$ transition. \\ \\
{\bf Fig. 6} The same as in Fig. 5, but for the helicity amplitude
$S_{1/2}$. \\ \\
{\bf Fig. 7} The same as in Fig. 1, but for the $\gamma^\ast \Xi^- \to
\Xi^{-\ast}$ transition, at $M^2=1.8~\mbox{GeV}^2$,
$s_0=4.0~\mbox{GeV}^2$. \\ \\
{\bf Fig. 8} The same as in Fig. 7, but for the helicity amplitude
$S_{1/2}$. \\ \\
{\bf Fig. 9} The same as in Fig. 7, but for the $\gamma^\ast \Xi^0 \to
\Sigma^{-\ast}$ transition. \\ \\
{\bf Fig. 10} The same as in Fig. 9, but for the helicity amplitude
$S_{1/2}$.

\newpage

\begin{figure}
\vskip 3. cm
    \includegraphics{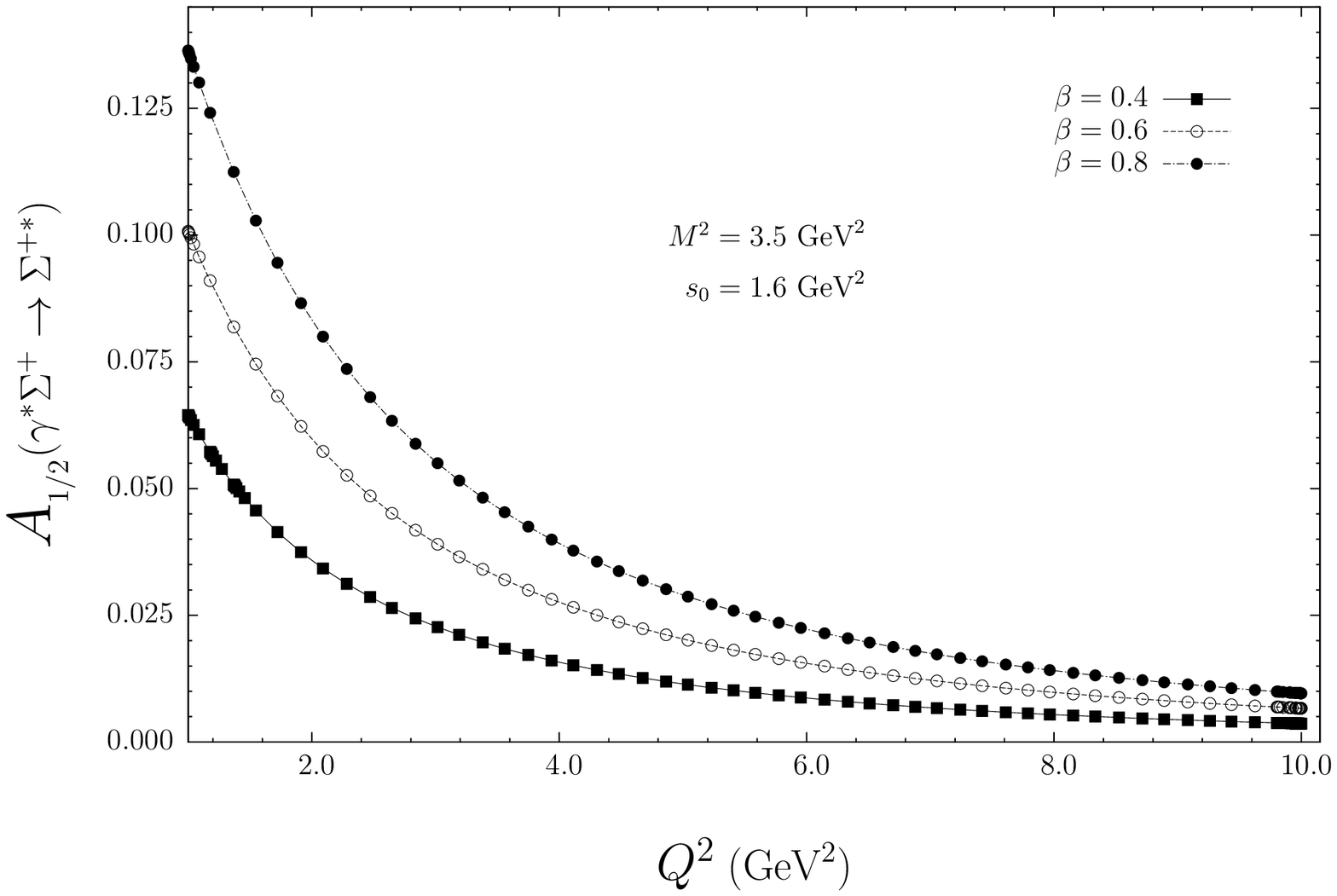}
\vskip 7.0cm
\caption{}
%\begin{center}
%{\bf Fig. 1-a}
%\end{center}
\end{figure}

\begin{figure}
\vskip 3. cm
    \includegraphics{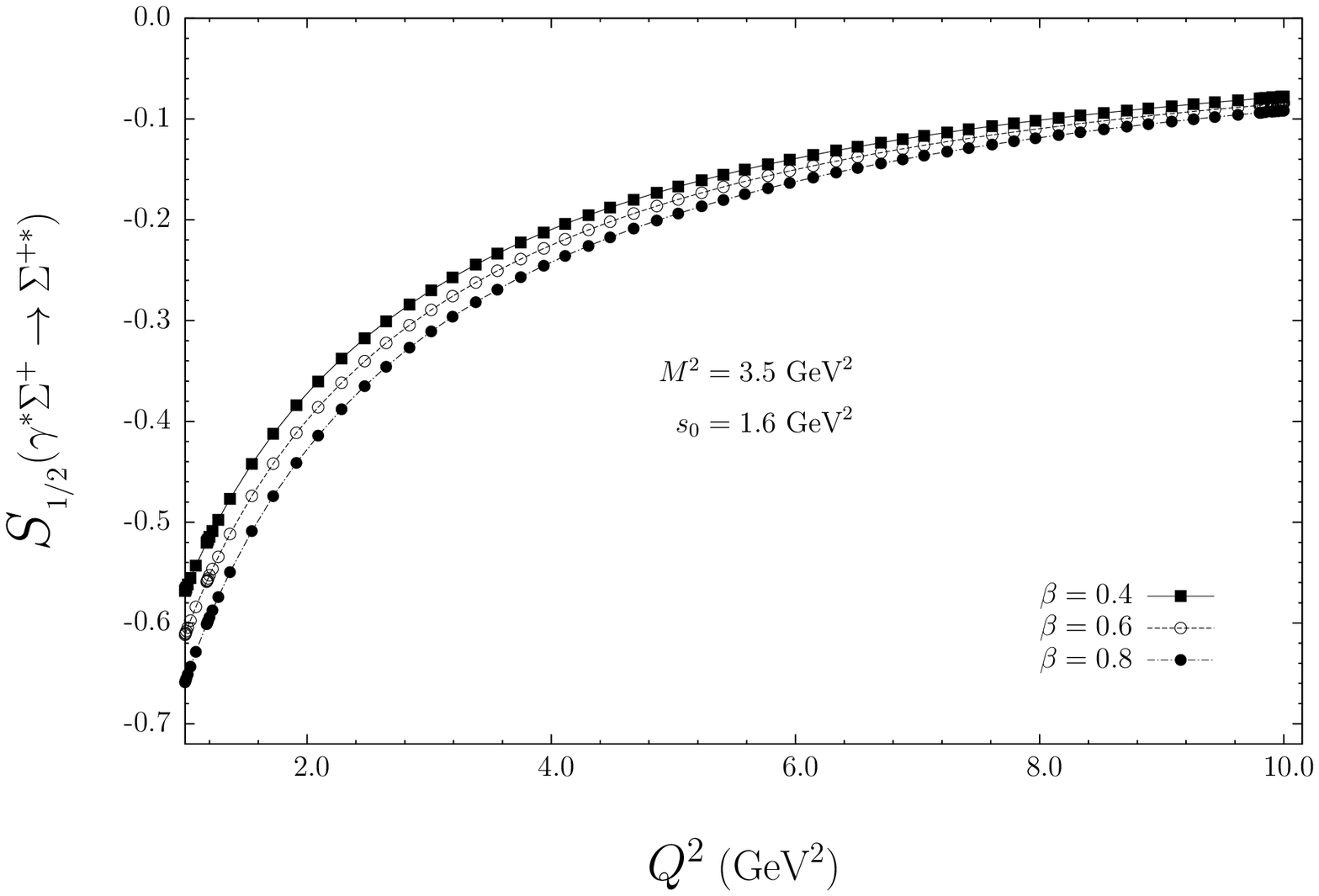}
\vskip 7.0cm
\caption{}
%\begin{center}
%{\bf Fig. 1-a}
%\end{center}
\end{figure}

\begin{figure}
\vskip 3. cm
    \includegraphics{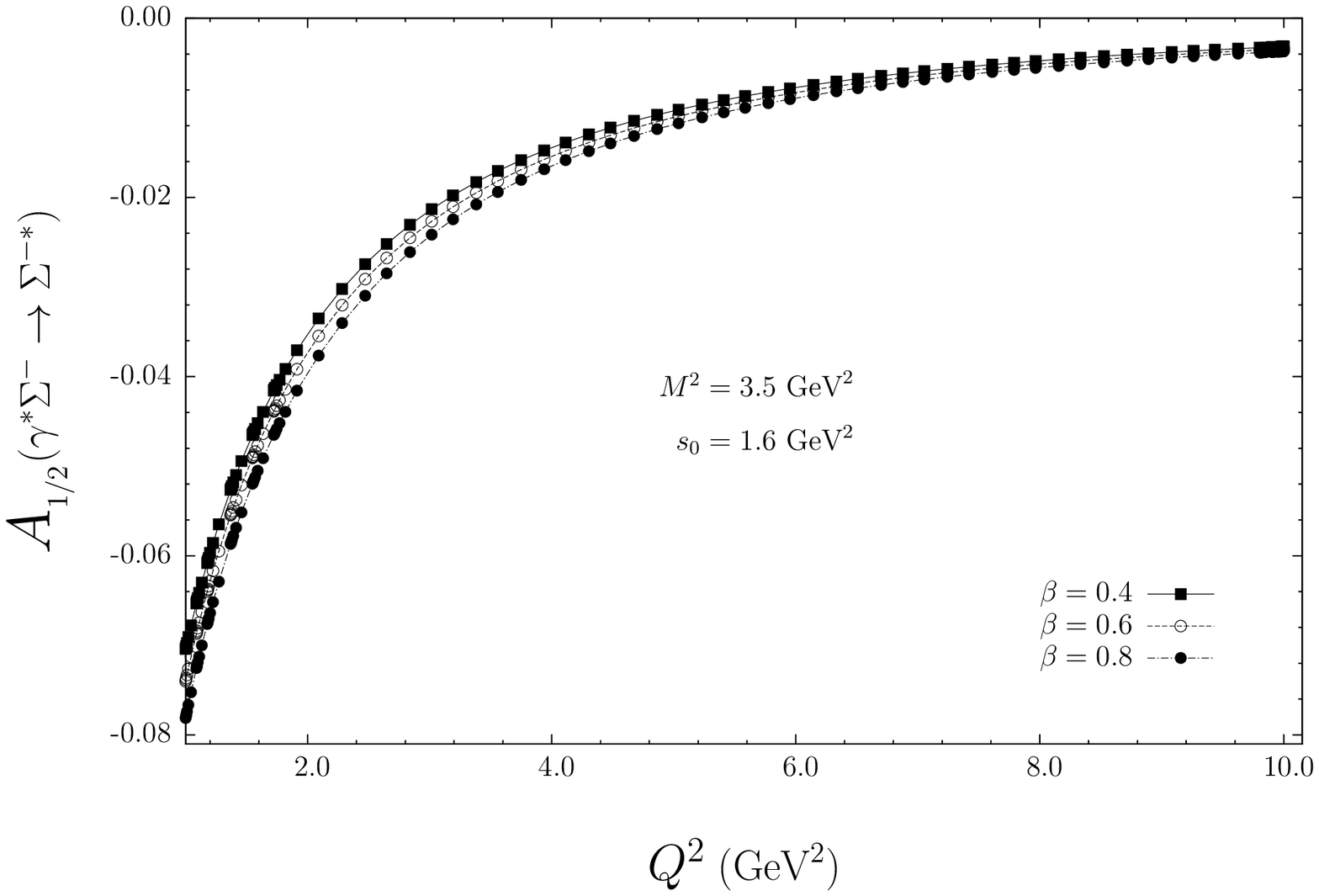}
\vskip 7.0cm
\caption{}
%\begin{center}
%{\bf Fig. 1-a}
%\end{center}
\end{figure}

\begin{figure}
\vskip 3. cm
    \includegraphics{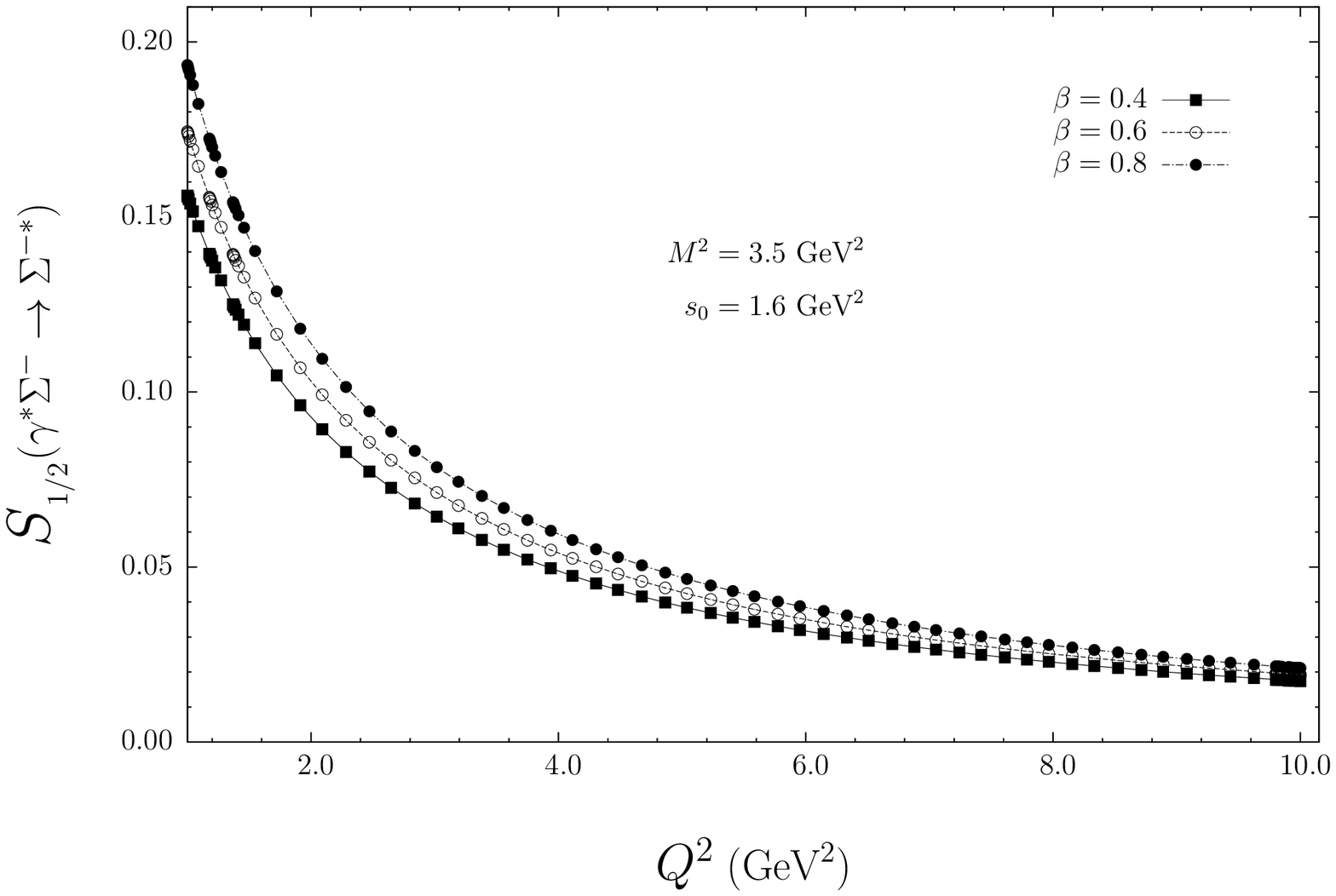}
\vskip 7.0cm
\caption{}
%\begin{center}
%{\bf Fig. 1-a}
%\end{center}
\end{figure}

\begin{figure}
\vskip 3. cm
    \includegraphics{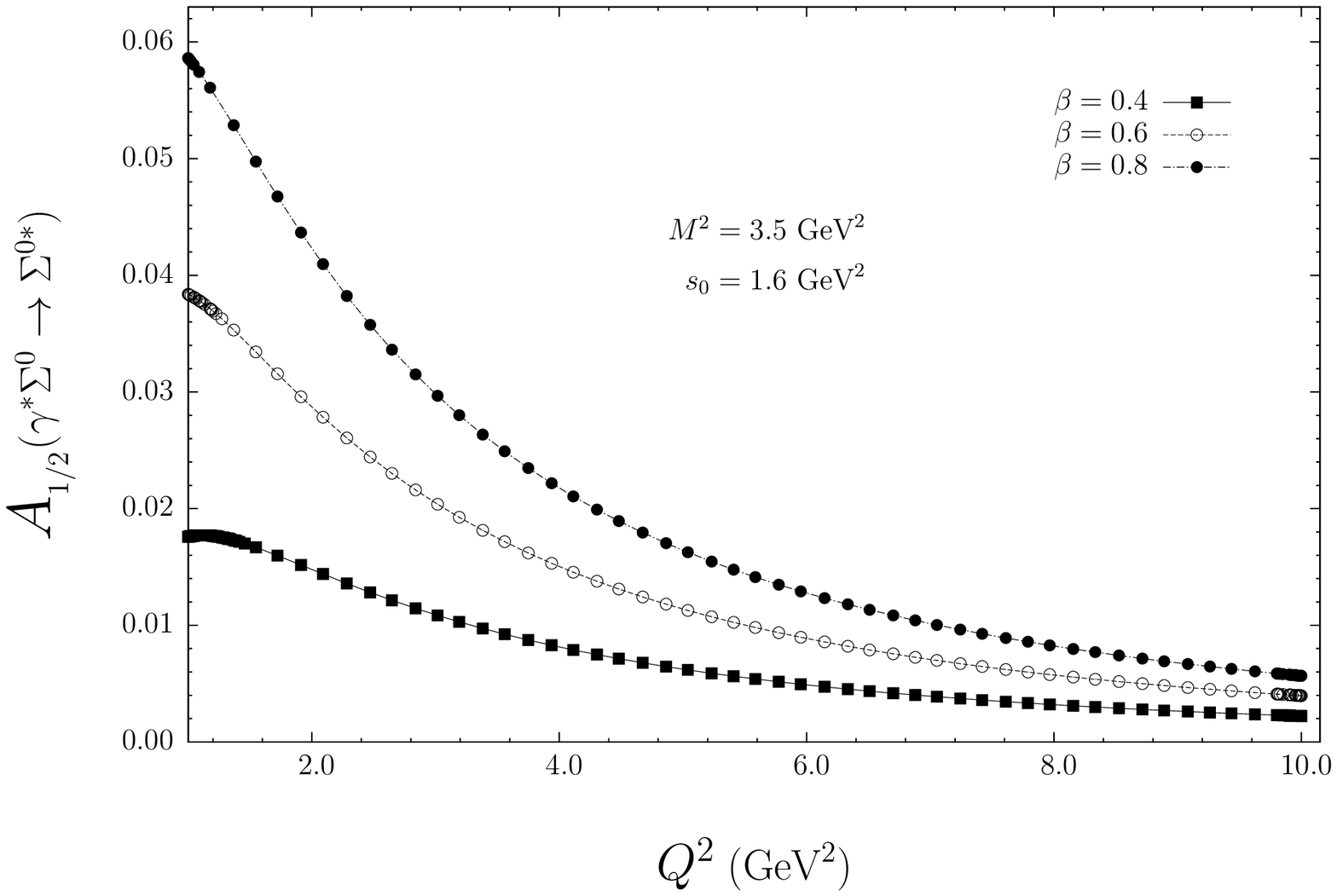}
\vskip 7.0cm
\caption{}
%\begin{center}
%{\bf Fig. 1-a}
%\end{center}
\end{figure}

\begin{figure}
\vskip 3. cm
    \includegraphics{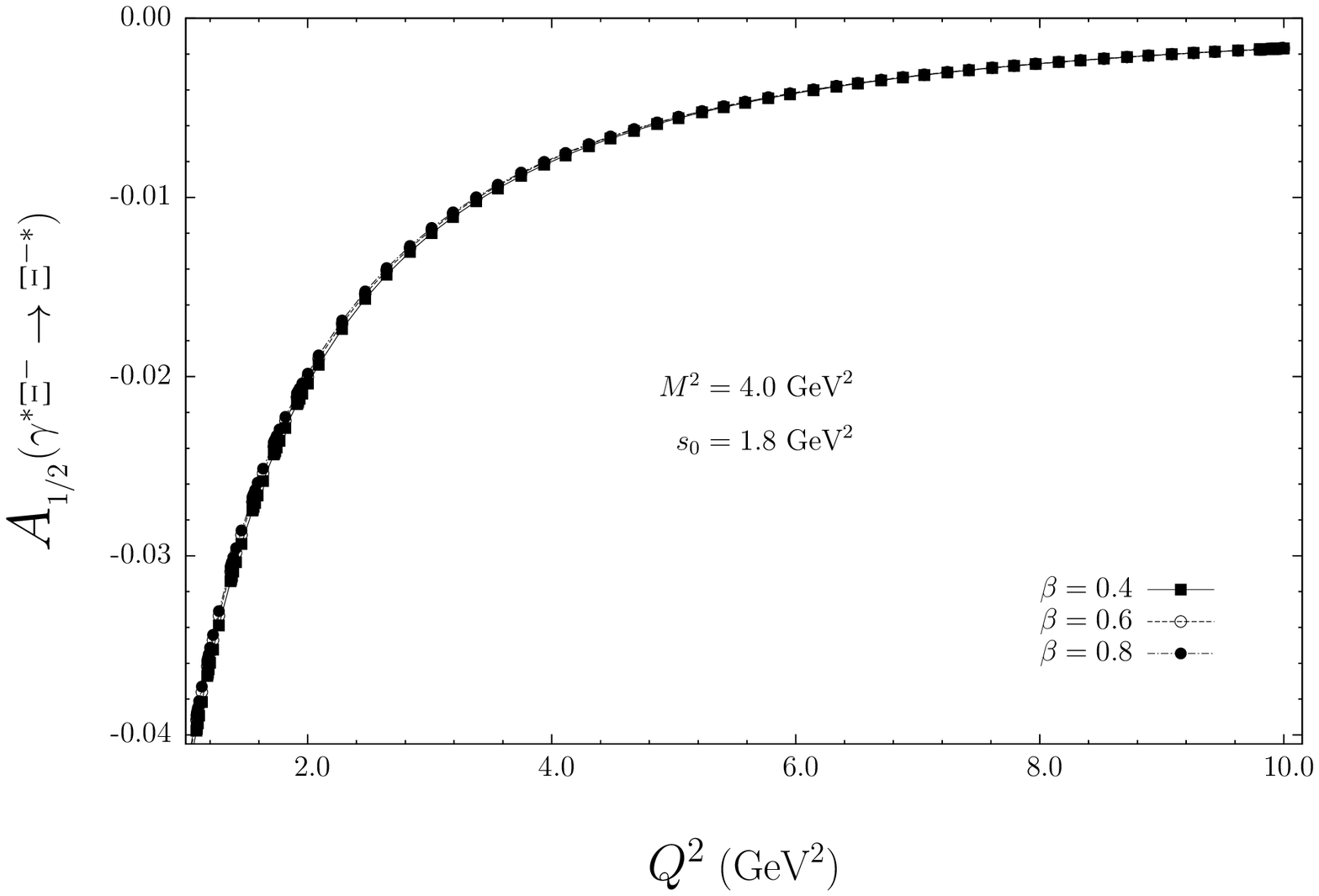}
\vskip 7.0cm
\caption{}
%\begin{center}
%{\bf Fig. 1-a}
%\end{center}
\end{figure}

\begin{figure}
\vskip 3. cm
    \includegraphics{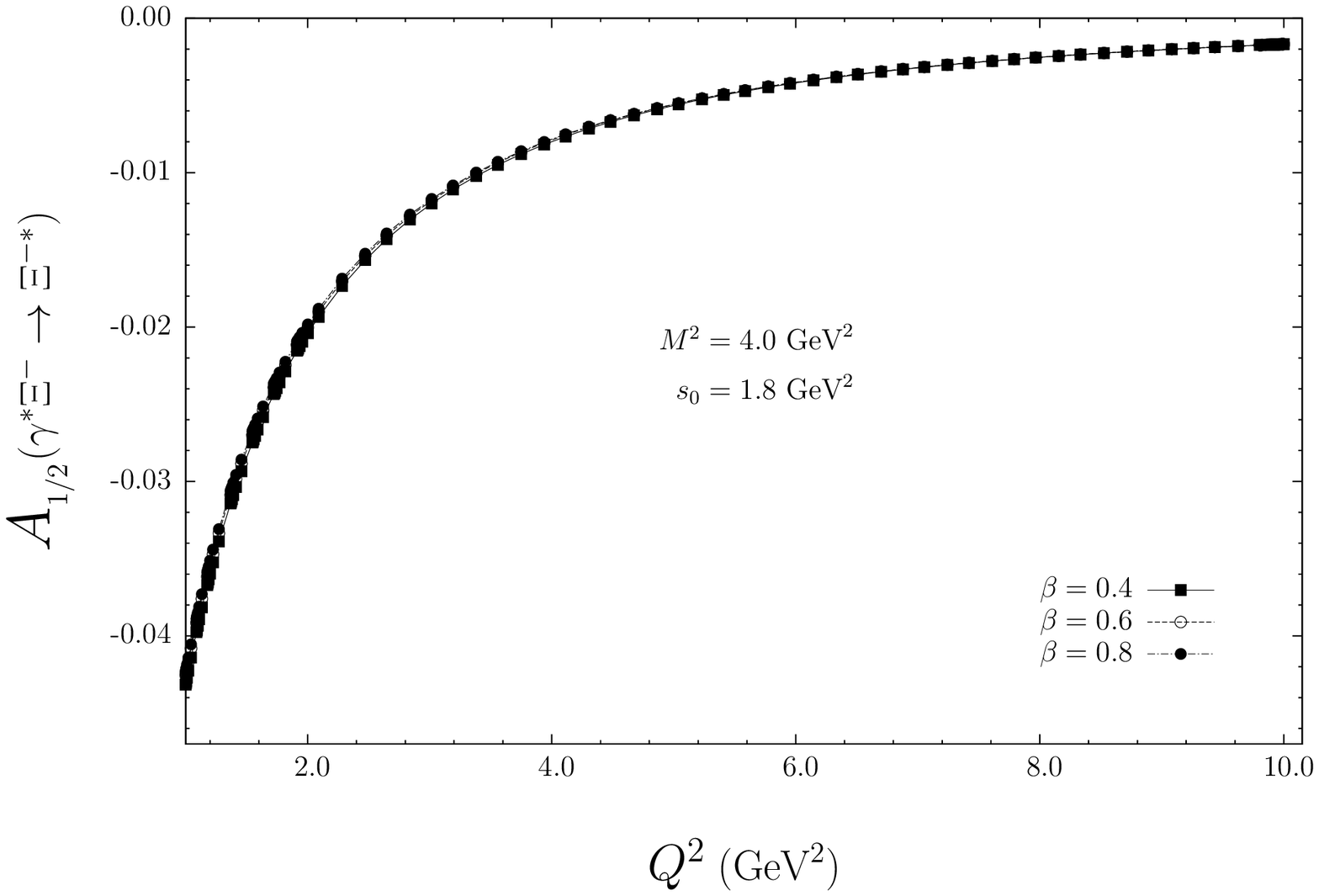}
\vskip 7.0cm
\caption{}
%\begin{center}
%{\bf Fig. 1-a}
%\end{center}
\end{figure}

\begin{figure}
\vskip 3. cm
    \includegraphics{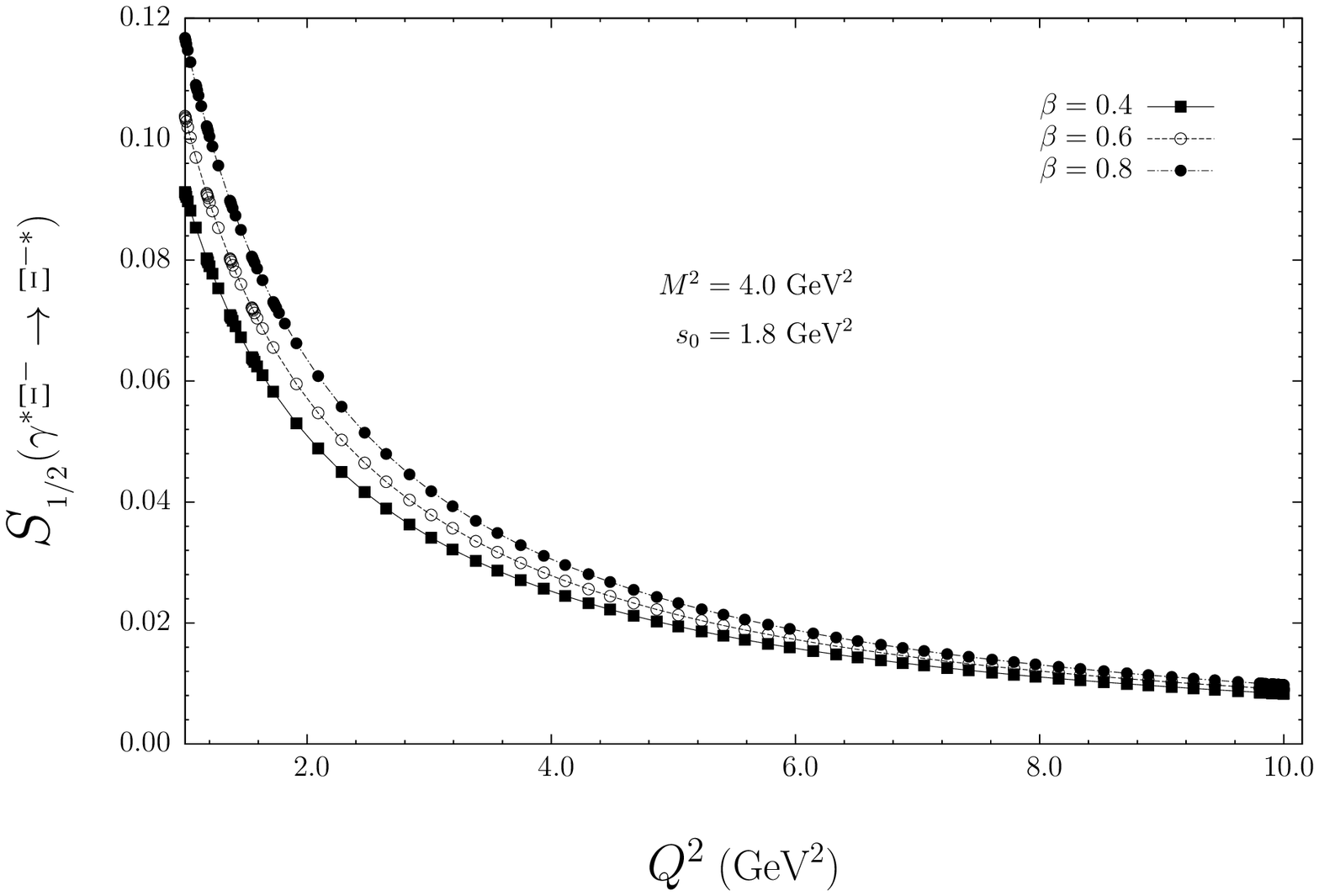}
\vskip 7.0cm
\caption{}
%\begin{center}
%{\bf Fig. 1-a}
%\end{center}
\end{figure}

\begin{figure}
\vskip 3. cm
    \includegraphics{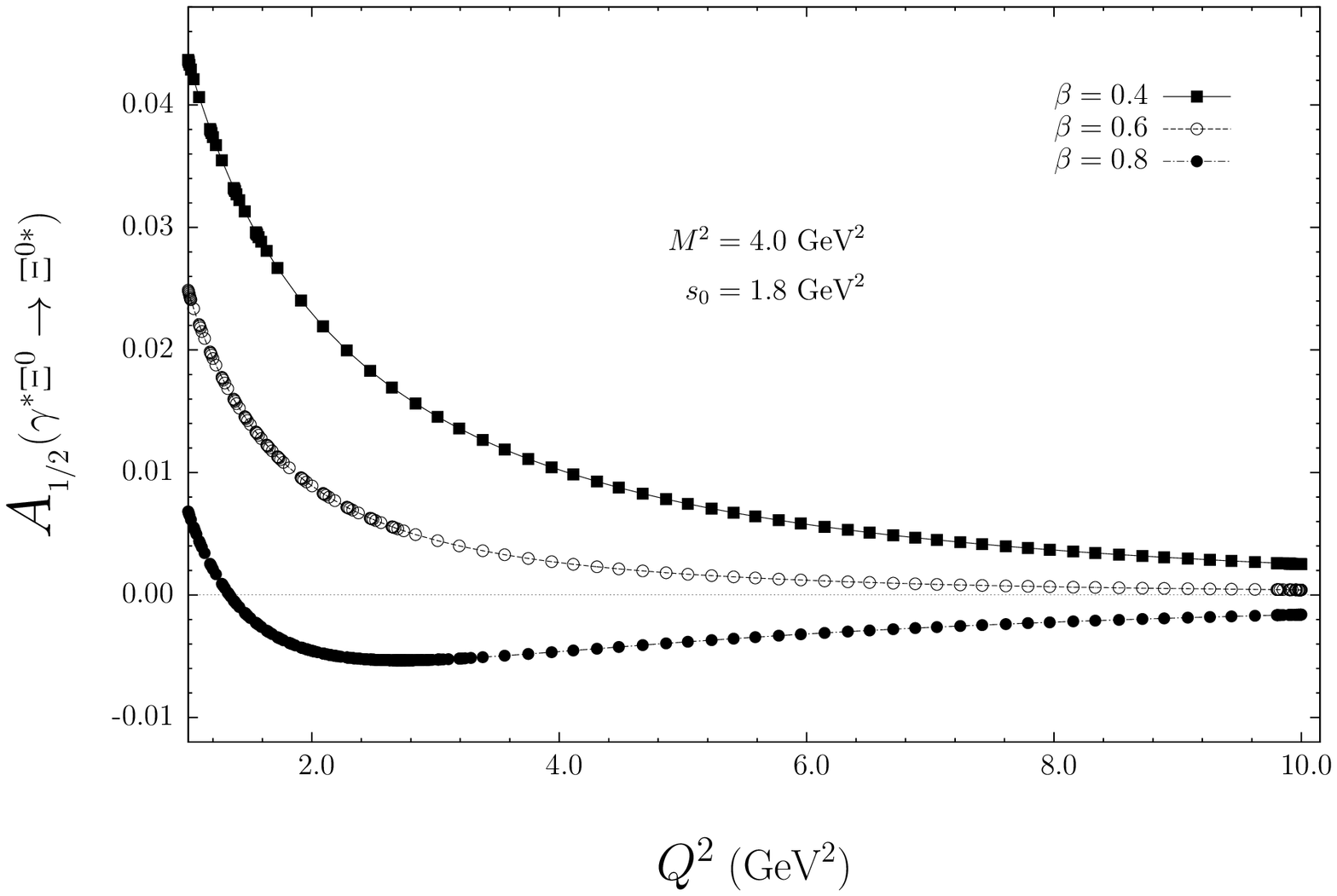}
\vskip 7.0cm
\caption{}
%\begin{center}
%{\bf Fig. 1-a}
%\end{center}
\end{figure}

\begin{figure}
\vskip 3. cm
    \includegraphics{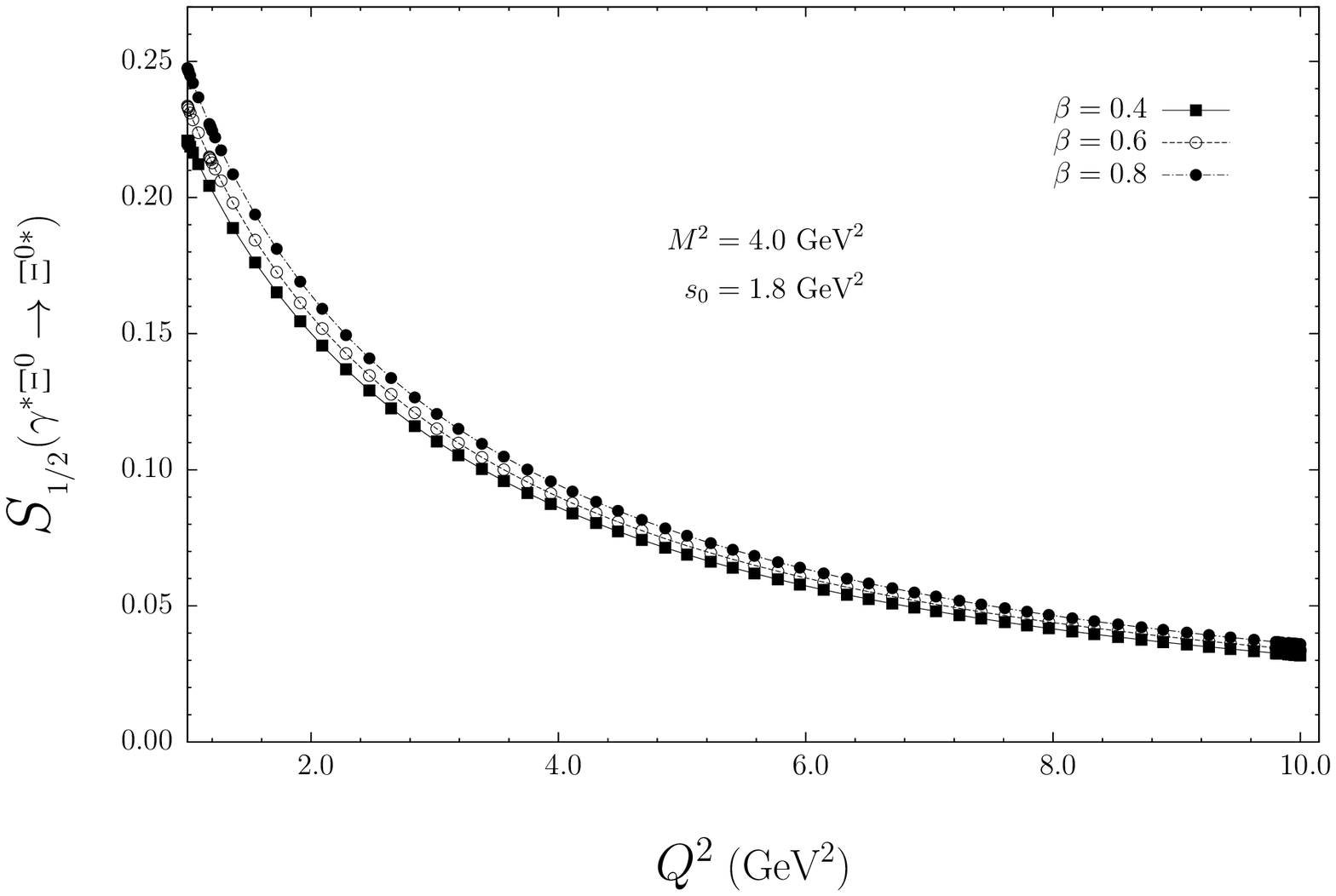}
\vskip 7.0cm
\caption{}
%\begin{center}
%{\bf Fig. 1-a}
%\end{center}
\end{figure}

\end{document}